\documentclass[journal,10pt,twocolumn,nofonttune]{IEEEtran}
\usepackage{xmpaper,lzqsymbol}
\usepackage{amssymb,amsmath,latexsym}
\usepackage{graphicx,subfigure}
\usepackage{subfigure}
\usepackage{color,cite,times}
\usepackage{epstopdf}
\usepackage{multirow}
\usepackage{array}
\usepackage{algorithm}
\usepackage{algorithmic}
\usepackage{booktabs}
\usepackage{bbm} 
\usepackage{subfigure}

\newcommand{\nc}{\newcommand}
\nc{\bsm}{\boldsymbol}
\nc{\mbs}{\mathbbmss}

\begin{document}
\title{Passive Beamforming and Information Transfer Design for Reconfigurable Intelligent Surfaces Aided Multiuser MIMO Systems}

\author{Wenjing Yan,
	Xiaojun~Yuan,~\IEEEmembership{Senior~Member,~IEEE}, Zhen-Qing He, and Xiaoyan Kuai

}
\maketitle

\begin{abstract}
  This paper investigates the passive beamforming and information transfer (PBIT) technique for the multiuser multiple-input multiple-output (MIMO) systems with the aid of a reconfigurable intelligent surface (RIS), where the RIS enhances the primary communication via passive beamforming and at the same time delivers additional information by the spatial modulation (which adjusts the on-off states of the reflecting elements). For the passive beamforming design, we propose to maximize the sum channel capacity of the RIS-aided multiuser MIMO channel and formulate the problem as a two-step stochastic program. A sample average approximation (SAA) based iterative algorithm is developed for the efficient passive beamforming design of the considered scheme. To strike a balance between complexity and performance, we then propose a simplified beamforming algorithm by approximating the stochastic program as a deterministic alternating optimization problem. For the receiver design, the signal detection at the receiver is a bilinear estimation problem since the RIS information is multiplicatively modulated onto the reflected signals of the reflecting elements. To solve this bilinear estimation problem, we develop a turbo message passing (TMP) algorithm in which the factor graph associated with the problem is divided into two modules: one for the estimation of the user signals and the other for the estimation of the RIS's on-off states. The two modules are executed iteratively to yield a near-optimal low-complexity solution. Furthermore, we extend the design of the multiuser MIMO PBIT scheme from single-RIS to multi-RIS, by leveraging the similarity between the single-RIS and multi-RIS system models. Extensive simulation results are provided to demonstrate the advantages of our passive beamforming and receiver designs.
\end{abstract}
\section{Introduction}
\label{sec.intro}

Reconfigurable intelligent surfaces (RISs), emerged as a new hardware technology to reduce the energy consumption and improve the spectrum efficiency of wireless networks, have recently attracted intensive research interest \cite{hum2014reconfigurable,di2019smart,basar2019wireless,nadeem2019large}.
 A RIS is an electromagnetic two-dimensional surface, composed of a large number of low-cost nearly-passive reconfigurable reflecting elements \cite{wu2019intelligent,huang2019reconfigurable}. As a prominent feature, the RIS can be flexibly implemented in practical communication scenarios (no matter the outdoor by installing it onto the facades of buildings, or the indoor by installing it onto the ceilings and walls of rooms) \cite{subrt2012intelligent}. Equipped with a smart controller, the RIS is able to intelligently adjust the phases of incident electromagnetic waves to increase the received signal energy, expand the coverage region, and alleviate interference, so as to enhance the communication quality of the wireless networks \cite{basar2019wireless,qingqing2019towards}. Besides, the RIS can also deliver additional information by adopting the spatial modulation on the index of the reflecting elements \cite{di2019smart,yan2019passive}.

In fact, traditional passive reflecting surfaces, which reflect electromagnetic waves with a fixed phase, have been extensively applied in radar and satellite communications for a long time \cite{hum2014reconfigurable}. However, the utilization of passive reflecting surfaces in terrestrial wireless communications had rarely been considered due to the time-varying environment of terrestrial communications. Thanks to the very recent development of metamaterials, it is now possible to reconfigure the phase shifters of the passive reflecting surfaces in real-time and thus to deploy RISs in terrestrial wireless networks \cite{foo2017liquid}. The RIS is advantageous over the existing related technologies in many aspects, such as multi-antenna relaying \cite{yuan2013multiple}, active intelligent surfaces \cite{hu2018beyond}, and backscattering \cite{zhang2019constellation}. For example, compared to multi-antenna relays and active intelligent surfaces, a passive RIS does not consume any energy in processing or retransmitting radio frequency signals; compared to backscatters, a RIS is able to conduct passive beamforming, i.e., to judiciously adjust the phases of the reflected electromagnetic waves to focus energy in the desired spatial directions, so as to significantly improve the energy-efficiency of wireless networks.

The incorporation of RISs into wireless networks poses a number of unprecedented challenges to the transceiver and RIS design \cite{basar2019wireless,nadeem2019large,qingqing2019towards}. For example, the authors in \cite{wu2019intelligent} studied the joint active and passive beamforming design to minimize the total transmit power, where both the active beamforming matrix at the transmitter and the passive beamforming matrix at the RIS are taken into account in the optimization. In \cite{huang2019reconfigurable}, transmit power allocation and passive beamforming were jointly designed to maximize energy/spectral efficiency. The optimization of passive beamforming requires the knowledge of the channel state information (CSI) of both the transmitter-RIS link and the RIS-receiver link. The CSI acquisition in a RIS-aided wireless communication network is a particularly challenging task due to the limited processing capability of the RIS. In this regard, the authors in \cite{taha2019enabling} assumed that a small portion of the RIS elements are ``active'' and are able to conduct baseband signal processing. Then, a channel estimation approach based on compressive sensing and deep learning was proposed for a RIS-aided multiple-input multiple-output (MIMO) channel. The channel estimation problem for a fully passive RIS was first tackled in \cite{he2019cascaded}, where a two-stage algorithm was developed to estimate the RIS-aided MIMO channel by utilizing sparse matrix factorization and matrix completion techniques.

Recently, a new RIS-aided wireless communication scheme, termed passive beamforming and information transfer (PBIT), was proposed to require that the RIS simultaneously enhances the primary communication (via passive beamforming) and delivers additional information to the receiver in a passive manner (by adopting the spatial modulation on the index of the reflecting elements) \cite{yan2019passive}. There are many potential sources of the RIS data.
For example, the RIS installed on a smart building is required to upload environmental data to the wireless network; the wireless control link (that coordinates the transmitter, the RIS, and the receiver for synchronisation and packed delivery) requires the RIS to acknowledge its status; or the CSI measured at the RIS (e.g., following the approach in \cite{taha2019enabling}) is required to be uploaded to a control center for assisting global resource allocation.
In the PBIT scheme, the reflecting coefficients of the RIS contain randomness since they carry the additional information delivered by the RIS, and so the passive beamforming design for a PBIT scheme generally involves difficult stochastic optimization. To simplify the problem, the authors in \cite{yan2019passive} proposed to maximize a heuristic performance metric, i.e., the average receive signal-to-noise ratio (SNR), and a semidefinite relaxation method was developed for the passive beamforming design of the single-input multiple-output (SIMO) PBIT scheme, where a single-antenna user communicates with a multi-antenna base station (BS) with the help of a RIS. Furthermore, the BS receiver of the SIMO PBIT scheme is required to reliably recover the information from both the user and the RIS, which gives rise to a bilinear detection problem. Efficient detection algorithms were developed in \cite{yan2019passive} by exploiting the rank-$1$ property of the received signal matrix.

In this paper, we study the design of the PBIT scheme for a multiuser MIMO system, where a number of single-antenna users communicate with a multi-antenna base station (BS) via the help of a RIS. Due to the multiplexing effect of the multiuser MIMO system, it is not appropriate to characterize the system performance by using a single SNR, and so the beamforming design in \cite{yan2019passive} is no longer applicable to this new scenario. Instead, we approximate the sum channel capacity by the conditional mutual information conditioned on the on-off states of the RIS elements, by noting that the information rate of the RIS is typically much lower than those of the users. Then, we propose to maximize the conditional mutual information and formulate the problem as a two-step stochastic program. A sample average approximation (SAA) based iterative algorithm is developed for the efficient passive beamforming design of the multiuser MIMO PBIT scheme. However, as a common issue of stochastic programming, the convergence speed of the SAA based beamforming algorithm is relatively slow. To strike a balance between complexity and performance, we further propose a simplified beamforming algorithm by approximating the stochastic program as a deterministic alternating optimization problem.

We also investigate the receiver design for the multiuser MIMO PBIT scheme. The receiver aims to retrieve the information from both the users and the RIS, which is a bilinear detection problem. Again, due to the multiplexing effect of the multiuser MIMO system, the received signal matrix at the BS is not rank-one, and therefore the rank-$1$ matrix factorization techniques developed in \cite{yan2019passive} are no longer applicable. Message passing is a powerful technique to yield near-optimal low complexity solutions to sophisticated inference problems \cite{gropp1999using}. Particularly, the parametric bilinear generalized approximate message passing (PBiGAMP) algorithm \cite{parker2016parametric} is designed to handle bilinear inference problems such as the one considered in this work. However, the complexity of the PBiGAMP algorithm is quadratic to the transmission block length, which is unaffordable when the large block length is large. More importantly, we observe that the PBiGAMP algorithm performs very poor in our scheme. The reason is that the PBiGAMP algorithm is specifically designed for measurement matrices composed of independent and identically distributed (i.i.d.) Gaussian entries, whereas the measurement matrixes for our problem are low-rank matrices far from i.i.d. Gaussian. To address these two issues, we develop a turbo message passing (TMP) algorithm to retrieve the information from both the users and the RIS. Specifically, we represent the inference problem by a factor graph and divide the whole factor graph into two modules, namely, one for the estimation of the user signals and the other for the estimation of the RIS's on-off states (that carry the RIS information). The estimation problem involved in each module is linear and hence can be solved by an existing algorithm named damped Gaussian generalized approximate message passing with sparse Bayesian learning (GGAMP-SBL)  \cite{al2017gamp}. The two modules are executed iteratively until convergence, hence the name turbo message passing. We show that the complexity of the proposed TMP algorithm is linear to the block length. We also show by numerical results that the TMP algorithm significantly outperforms the PBiGAMP algorithm, and is able to closely approach the performance lower bounds (obtained by assuming either known user data or known RIS data).

Furthermore, we extend the design of the multiuser MIMO PBIT scheme from single-RIS to multi-RIS, where a number of RISs are installed in different locations to cooperatively enhance the user-BS communications as well as the RIS-BS information transfer. To the best of our knowledge, this is the first work to study the passive beamforming design and the receiver design for the multi-RIS scenario. We first establish the similarity between the multi-RIS system model and the single-RIS system model. Based on the model similarity, we show that both the passive beamforming and receiver designs for the single-RIS case straightforwardly carry over to the multi-RIS case, except for some minor modifications to the prior distribution of the on-off states of the RIS elements.


{\it Notation:}
For any matrix $\bsm{A}$, $\bsm{a}_i$ refers to the $i$th column of $\bsm{A}$, and $a_{ij}$ refers to the $(i,j)$th entry of $\bsm{A}$. $\mathbb{C}$ denotes the complex field; $\mathbb{R}$ denotes the real field; $\mathcal{S}$ denotes a set, and $|\mathcal{S}|$ represents the the cardinality of $\mathcal{S}$. $|x|$ represents the absolute value of $x$; $\|\cdot\|_2$ represents the $\ell_2$-norm; $\|\cdot\|_{\rm F}$ represents the Frobenius norm.
The superscripts $(\cdot)^{\rm T}$, $(\cdot)^{\ast}$, $(\cdot)^{\rm H}$, $(\cdot)^{-1}$ represent the  transpose, the conjugate, the conjugate transpose, and the inverse of a matrix, respectively. $\odot$ represents the Hadamard product; $\otimes$ represents the Kronecker product.
 $\mbs{E}(\cdot)$ and $\mbs{Var}(\cdot)$ represent the expectation and the variance, respectively.
 $\delta(\cdot)$  represents the Dirac delta function.
diag$\{\bsm{a}\}$ represents the diagonal matrix with the diagonal specified by $\bsm{a}$;  diag$\{\bsm{A}\}$ represents the vector composed of the diagonal elements of those of $\bsm{A}$;
$\tr(\bsm{A})$ and $\det(\bsm{A})$ represent the trace and the determinant of square matrix $\bsm{A}$, respectively.
For any integer $N$,  $\mathcal{I}_N$ denotes the set of integers from $1$ to $N$; $\bsm{I}$ is the identity matrix with an appropriate size;  $\textbf{1}_N$ represents the $N$-dimensional all-one vector.
$\mathcal{CN}(\cdot;\mu,\nu)$ represents a complex Gaussian distribution with mean $\mu$, covariance $\nu$, and relation zero.
$\text{vec}(\bsm{A})$ represents the vector obtained by stacking the columns of the matrix $\bsm{A}$ sequentially.

\section{System Model and Problem Description} \label{sec.model}
\subsection{System Model} \label{sec.sys}
\begin{figure}
  \centering
  \includegraphics[width=2.8 in]{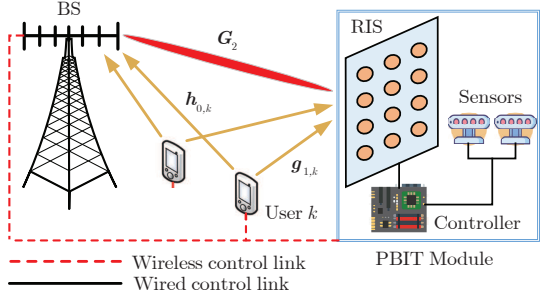}\\
  \caption{A PBIT-enhanced uplink MIMO system with a RIS.}\label{RIS}
\end{figure}

As illustrated in Fig.~\ref{RIS}, we consider a PBIT-enhanced uplink MIMO multiuser system, where $K$ single-antenna users communicate with a base station (BS) equipped with $M$ antennas via the help of a RIS. Beyond traditional end-to-end wireless communications, the system adds an additional PBIT module to simultaneously enhance the active user-BS communication and achieve the passive RIS-BS communication.
The PBIT module is an intelligent reflecting device that consists of a RIS equipped with $N$ passive reflecting elements, a controller to adaptively adjust the on/off states and the phase shifts of the passive reflecting elements, and a number of sensors or other Internet of Things (IoT) devices that collect the environmental data. The sensors are connected to the RIS by wires.
To enhance the active user-BS communication, the RIS reflects the incident signals transmitted from the users by the activated reflecting elements. Further, the phases of the reflected signals can be adjusted by the controller to optimize the user-BS communication performance.
To achieve the passive RIS-BS communication, the controller adjusts the on/off state of each reflecting element according to the wired data from the sensors (i.e., the on/off states of the reflecting elements carry the sensors' information).

Denote by $\bsm{H}_{0}=[\bsm{h}_{0,1},\ldots,\bsm{h}_{0,K}]\in \mathbb{C}^{M \times K}$
 the baseband equivalent channel of the user-BS link, where $\bsm{h}_{0,k}\in \mathbb{C}^{M \times 1} $ is the channel coefficient vector of user $k$.
 Denote by $\bsm{G}_{1}=[\bsm{g}_{1,1},\ldots,\bsm{g}_{1,N}]^{\rm T }\in \mathbb{C}^{N \times K}$
 the baseband equivalent channel of the user-RIS link, where $\bsm{g}_{1,n}\in \mathbb{C}^{K \times 1} $ is the channel coefficient vector between all users and the $n$th reflecting element of the RIS.
 Let $s_n$ be the state of the $n$th passive reflecting element, where $s_n$ takes the value $0$ or $1$ to represent the ``on'' or ``off'' state of the $i$th element. $\bsm{S} = \text{diag}\{ \bsm{s}\}$ is the diagonal state matrix of the RIS, where
 $\bsm{s} = [s_1, s_2, \ldots, s_N]^{\rm T}$ carries the information from the sensors.
  Let $\varphi_n$ be the phase shift of the $n$th passive reflecting element, where $\varphi_n\in [0,2\pi]$. Then $\bsm{\Theta} = \text{diag}\{ \bsm{\theta}\}$ is the diagonal phase-shift matrix of the RIS, where
 $ \bsm{\theta} = [\theta_1, \theta_2, \ldots, \theta_N]^{\rm T}$ and $\theta_n = e^{j\varphi_n}$.
  Denote by $\bsm{G}_{2}=[\bsm{g}_{2,1},\ldots,\bsm{g}_{2,N}]\in \mathbb{C}^{M \times N}$ the baseband equivalent channel of the RIS-BS link, where  $\bsm{g}_{2,n}\in \mathbb{C}^{M \times 1} $ is the channel coefficient vector between the $n$th reflecting element of the RIS and the BS.
We neglect the signal power reflected by the RIS for two or more times due to severe path loss.

Assume that the channel is block-fading with each transmission block consisting of $T$ time slots.  The observed signal at the BS
in time slot $t$ is
\begin{align} \label{channel.H2}
&\bsm{y}_t = ( \bsm{G}_2 \bsm{\Theta}\bsm{S}\bsm{G}_1  + \bsm{H}_0 )\bsm{x}_t +
 \bsm{w}_t,
\end{align}
where $\bsm{x}_t\in \mathbb{C}^{K \times 1}$ is the transmit signal vector in time slot $t$
and $\bsm{w}_t\in \mathbb{C}^{M \times 1}$ is an additive white Gaussian noise (AWGN) with the elements independently drawn from $\mathcal{CN}(0, \sigma_w^2)$.
  Assume that the diagonal state matrix $\bsm{S}$ and the phase-shift matrix $\bsm{\Theta}$ of the RIS remain fixed over a transmission block.  Then, the observed signal matrix in a transmission block, denoted by
$\bsm{Y}=[\bsm{y}_1, \ldots, \bsm{y}_T]$, can be expressed as
\begin{align} \label{channel.H1}
&\bsm{Y} = ( \bsm{G}_2 \bsm{\Theta}\bsm{S}\bsm{G}_1  + \bsm{H}_0 )\bsm{X} +
 \bsm{W} ,
\end{align}
where $\bsm{X}\triangleq[\bsm{x}_1,\ldots,\bsm{x}_T]$ and $\bsm{W}\triangleq[\bsm{w}_1, \ldots, \bsm{w}_T]$.

Each entry of $\bsm{X}$ is independently modulated by using a constellation $\mathcal{C}=\{c_1,c_2,\ldots,c_{\vert \mathcal{C}\vert}\}$. Then, the probability density function (PDF) of $\bsm{X}$ is given by
\begin{align} \label{pro.x}
p_{\bsm{X}}(\bsm{X})= \prod_{k=1}^K \prod_{t=1}^T p_{x_{kt}}(x_{kt}) = \prod_{k=1}^K \prod_{t=1}^T \frac{1}{|\mathcal{C}|} \sum_{i=1}^{|\mathcal{C}|} \delta( x_{kt}- c_i).
\end{align}
The transmitted signals of each user $k$ in a transmission block is power-constrained as
\begin{align}
\frac{1}{T} \sum_{t=1}^T \mbs{E}|x_{kt}|^2 \leqslant p_k, \quad \forall k\in \mathcal{I}_K ,
\end{align}
where $p_k$ is the power budget of user $k$.

Spatial modulation is applied by manipulating the on-off states of the reflecting elements. Specifically, we assume that $s_n,\forall n \in \mathcal{I}_{N}$, independently takes the value of $1$ (meaning that the state of the $n$th RIS element is ``on'') with probability $\rho$ and the value of $0$ (meaning that the state of the $n$th RIS element is ``off'') with probability $1-\rho$, yielding
\begin{align} \label{pro.s}
  p(\bsm{s}) = \prod_{n=1}^N p(s_n) = \prod_{n=1}^N\left( (1-\rho)\delta(s_n) + \rho \delta(s_n-1) \right).
\end{align}
Note that $\rho$ is also referred to as the sparsity of the RIS.
The information carried by each passive reflecting element is $H(\rho)=-\rho \log_2 \rho - (1-\rho) \log_2 (1-\rho)$ bit.


\subsection{Problem Description}

In the PBIT-enhanced uplink MIMO system described above, the receiver is required to retrieve the information from both the users and the RIS (i.e., $\boldsymbol{X}$ and  $\boldsymbol{s}$) under the assumption of perfect channel state information (CSI).\footnote{The CSI acquisition techniques of the RIS-aid MIMO channel can be found, e.g., in \cite{taha2019enabling} and \cite{he2019cascaded}. }
Furthermore,
the phase shifts of the passive reflecting elements in the RIS need to be carefully adjusted to enhance the recovery performance at the receiver.
A natural design criterion is to maximize the sum channel capacity.
 From information theory, the sum channel capacity of the PBIT system is given by the mutual information $I(\bsm{x}_t,\bsm{s};\bsm{y}_t)$\footnote{Since $I(\bsm{x}_t,\bsm{s};\bsm{y}_t)$ is invariant to index $t$, we henceforth omit the index $t$ by simply writing $I(\bsm{x},\bsm{s};\bsm{y})$. }\cite{cover2012elements}.
Then, our design problem can be divided into two subproblems: One is the passive beamforming design, i.e., to maximize $I(\boldsymbol{x},\boldsymbol{s};\boldsymbol{y})$ over the phase shift matrix $\boldsymbol{\Theta}$; and the other is the transceiver design, i.e., to design the signaling of $\boldsymbol{x}$ and $\boldsymbol{s}$ at the users and the RIS as well as the receiver at the BS to achieve the maximized $I(\boldsymbol{x},\boldsymbol{s};\boldsymbol{y})$.

We first consider the passive beamforming design to maximize $I(\bsm{x},\bsm{s};\bsm{y})$ over $\bsm{\Theta}$. Note that $I(\boldsymbol{x},\boldsymbol{s};\boldsymbol{y})$ is in general difficult to evaluate since \eqref{channel.H2} is a complicated model involving the multiplication of $\bsm{s}$ and $\bsm{X}$. To avoid this difficulty, we propose an approximate design metric as follows.
Recall that the data of the RIS are collected from the sensors or other IoT devices. It is known that the data rate of an IoT device is typically much lower than that of a cellular user.  Thus, we have
\begin{align}
&I(\bsm{x},\bsm{s}; \bsm{y})= I(\bsm{x};\bsm{y}|\bsm{s}) + I(\bsm{s};\bsm{y})  \approx I(\bsm{x};\bsm{y}|\bsm{s}) \label{Inf.2},
\end{align}
where the first step follows from the chain rule of mutual information.
Then, the passive beamforming design problem is converted to maximizing $I(\bsm{x};\bsm{y}|\bsm{s})$ over $\bsm{\Theta}$.

For the transceiver design, we will focus on the design of the receiver at the BS to reliably recover both $\boldsymbol{X}$ and $\boldsymbol{s}$ from the received signal $\boldsymbol{Y}$ (for a given $\boldsymbol{\Theta}$). From information theory, besides the receiver design, we also need to design signal shaping and channel coding at the transmitter, so as to approach the channel capacity.  The signal shaping and channel coding design is, however, out of the scope of this paper.

\section{Beamforming Design} \label{sec.beamf1}

We now consider the beamforming design of $\bsm{\Theta}$ to maximize the conditional mutual information $I(\bsm{x};\bsm{y}|\bsm{s})$, where $\bsm{x}$, $\bsm{y}$ and $\bsm{s}$ are the corresponding variables in  \eqref{channel.H2} by ignoring the index $t$.
This problem can be represented by a two-stage stochastic program as
\begin{subequations} \label{optimal.e}
\begin{align}
 \min_{\bsm{\Theta}}\quad   &\mbs{E}_{\bsm{s}} \mathcal{Q}(\bsm{\Theta},\bsm{s} ) \label{optimal.e.a}\\
\textrm{s.t.} \quad~  &|\theta_n| = 1, \forall n \in \mathcal{I}_N , \label{optimal.e.b} \\
&\mathcal{Q}(\bsm{\Theta},\bsm{s} ) = \min_{\bsm{\Phi},\bsm{\Sigma}} \mbs{E}_{\bsm{y},\bsm{x}|\bsm{s}} \left( \log \det(\bsm{\Sigma}) \right. \notag\\
&~~~~~~~~~~~~~~~~~~\left. + \left\|\bsm{\Sigma}^{-\frac{1}{2}} \left(\bsm{x}- \bsm{\Phi}\bsm{y}\right)\right\|_2^2\right) \label{optimal.e.c}
\end{align}
\end{subequations}
where $\bsm{\Phi} \in \mathbb{C}^{K \times M}$ is a combining matrix, and $\bsm{\Sigma} \in \mathbb{C}^{K \times K} $ is positive semidefinite. The detailed formulation of the problem in  \eqref{optimal.e} can be found in Appendix A.
To calculate the expectation in \eqref{optimal.e.a}, we need to enumerate all the $2^N$ scenarios of $\bsm{s}$, which is difficult especially when $N$ is large.
Thus, we follow the SAA method \cite{kleywegt2002sample} to approximate \eqref{optimal.e} by independently generating $\ell_{\bsm{s}}$ replications $\{\bsm{s}_{[1]},\ldots,\bsm{s}_{[\ell_{\bsm{s}}]} \}$ of $\bsm{s}$ based on the probability model in \eqref{pro.s}. It has been shown that the solution of the SAA approximation approaches the optimal solution of the original stochastic program with probability one for a sufficiently large sample size \cite{dai2000convergence,shapiro2000rate}.
Then, the stochastic program in \eqref{optimal.e} is approximated by
\begin{subequations} \label{optimal.a1}
\begin{align}
 \min_{\bsm{\Theta}} \quad & \frac{1}{\ell_{\bsm{s}}} \sum_{i=1}^{\ell_{\bsm{s}}} \mathcal{Q}(\bsm{\Theta},\bsm{s}_{[i]} ) \label{optimal.a1.a}\\
\textrm{s.t.} \quad~  & |\theta_n| = 1, \forall n \in \mathcal{I}_N \label{optimal.a1.b} \\
&\mathcal{Q}(\bsm{\Theta},\bsm{s}_{[i]} ) = \min_{\bsm{\Phi},\bsm{\Sigma}} \mathcal{J}_{[i]}(\bsm{\Phi},\bsm{\Sigma}), \forall i \in \mathcal{I}_{\ell_{\bsm{s}}} \label{optimal.a1.c}
\end{align}
\end{subequations}
where
\begin{align} \label{optimal.J}
\mathcal{J}_{[i]}(\bsm{\Phi},\bsm{\Sigma}) &= \mbs{E}_{\bsm{y},\bsm{x}|\bsm{s}_{[i]}} \left( \log \det(\bsm{\Sigma}) \right. \notag\\
&~~~~ + \left. \left\|\bsm{\Sigma}^{-\frac{1}{2}} \left(\bsm{x}- \bsm{\Phi} \bsm{y}_{[i]} \right)\right\|_2^2\right), \forall i \in \mathcal{I}_{\ell_{\bsm{s}}}.
\end{align}
In the above,
$\bsm{y}_{[i]} = ( \bsm{G}_2 \bsm{\Theta}\bsm{S}_{[i]}\bsm{G}_1  + \bsm{H}_0 )\bsm{x} +  \bsm{w}$ is the replication of $\bsm{y}$ corresponding to $\bsm{s}_{[i]}$.
From \cite{dai2000convergence} and \cite{shapiro2000rate},
the solution of \eqref{optimal.a1} converges to the solution of \eqref{optimal.e} exponentially fast with the increase of the sample size $\ell_{\bsm{s}}$. Thus, a moderate $\ell_{\bsm{s}}$ is sufficient to find a relatively accurate solution of \eqref{optimal.e}.
In what follows, we propose a two-step alternating optimization method to solve the problem in \eqref{optimal.a1}: First optimize $\bsm{\Theta}$ for given $\{\mathcal{Q}(\bsm{\Theta},\bsm{s}_{[i]} )\}$, and then solve $\{\mathcal{Q}(\bsm{\Theta},\bsm{s}_{[i]} )\}$ for given $\bsm{\Theta}$.

\subsection{ Optimization of $\bsm{\Theta}$ for Given $\{\mathcal{Q}(\bsm{\Theta},\bsm{s}_{[i]} )\}$}
\label{sys.theta1}

For given $\{\mathcal{Q}(\bsm{\Theta},\bsm{s}_{[i]} )\}$, the problem in \eqref{optimal.a1} reduces to
\begin{subequations} \label{optimal.a2}
\begin{align}
 \min_{\bsm{\Theta}} \quad  &\frac{1}{\ell_{\bsm{s}}} \sum_{i=1}^{\ell_{\bsm{s}}} \mbs{E}_{\bsm{y},\bsm{x}|\bsm{s}_{[i]}} \left\|\bsm{\Sigma}^{-\frac{1}{2}}\left(\bsm{x}- \bsm{\Phi}\bsm{y}_{[i]} \right)\right\|_2^2 \\
~\textrm{s.t.} \quad~  &|\theta_n| = 1, \forall n \in \mathcal{I}_N.
\end{align}
\end{subequations}
From the model in \eqref{channel.H2}, we obtain
\begin{align} \label{optimal.a3}
& \mbs{E}_{\bsm{y},\bsm{x}|\bsm{s}_{[i]}} \left\|\bsm{\Sigma}^{-\frac{1}{2}}\left(\bsm{x}- \bsm{\Phi}\bsm{y}_{[i]} \right)\right\|_2^2 \notag\\
&= \mbs{E}_{\bsm{y},\bsm{x}|\bsm{s}_{[i]}} \left\|\bsm{\Sigma}^{-\frac{1}{2}}\left(\bsm{x}- \bsm{\Phi}((\bsm{G}_2 \bsm{\Theta}\bsm{S}_{[i]}\bsm{G}_1  + \bsm{H}_0 )\bsm{x} +
 \bsm{w})\right)\right\|_2^2  \notag\\
&= \tr \left\{ \bsm{\Sigma}^{-1} \left( \bsm{Q} - \bsm{\Phi} \bsm{H}_0 \bsm{Q}  - (\bsm{\Phi} \bsm{H}_0 \bsm{Q})^{\rm H} + \bsm{\Phi}\bsm{H}_0 \bsm{Q} \bsm{H}_0^{\rm H}\bsm{\Phi}^{\rm H}  \right.\right. \notag\\
&~~~~+ \sigma_{w}^2 \bsm{\Phi}\bsm{\Phi}^{\rm H} + \bsm{\Phi} \bsm{G}_2 \bsm{S}_{[i]}  \bsm{\Theta}\bsm{G}_1 \bsm{Q} \bsm{G}_1^{\rm H}  \bsm{\Theta}^{\rm H} \bsm{S}_{[i]}^{\rm H} \bsm{G}_2^{\rm H} \bsm{\Phi}^{\rm H} \notag\\
&~~~~+ \bsm{\Phi}\bsm{G}_2 \bsm{\Theta}\bsm{S}_{[i]} \bsm{G}_1 \bsm{Q} \bsm{H}_0^{\rm H}\bsm{\Phi}^{\rm H} + (\bsm{\Phi}\bsm{G}_2 \bsm{\Theta}\bsm{S}_{[i]} \bsm{G}_1 \bsm{Q} \bsm{H}_0^{\rm H}\bsm{\Phi}^{\rm H})^{\rm H}\notag\\
&~~~~ \left.\left. -\bsm{\Phi} \bsm{G}_2 \bsm{\Theta}\bsm{S}_{[i]} \bsm{G}_1 \bsm{Q} - (\bsm{\Phi} \bsm{G}_2 \bsm{\Theta}\bsm{S}_{[i]} \bsm{G}_1 \bsm{Q})^{\rm H} \right)\right\}.
\end{align}
To simplify the expression in \eqref{optimal.a3}, we decompose $\bsm{G}_1 \bsm{Q} \bsm{G}_1^{\rm H}$ as
\begin{align}
\bsm{G}_1 \bsm{Q} \bsm{G}_1^{\rm H} = \sum_{k=1}^{K} p_k \bsm{g}_{1,k} \bsm{g}_{1,k}^{\rm H}.  \label{optimal.a10}
\end{align}
Plugging \eqref{optimal.a10} into \eqref{optimal.a3}, we obtain
\begin{align}
&\min_{\bsm{\Theta}} \mbs{E}_{\bsm{y},\bsm{x}|\bsm{s}_{[i]}} \left\|\bsm{\Sigma}^{-\frac{1}{2}}\left(\bsm{x}- \bsm{\Phi}\bsm{y}_{[i]} \right)\right\|_2^2  \notag\\
&= \min_{\bsm{\Theta}} \sum_{k=1}^{K} p_k \bsm{\theta}^{\rm H} \diag\{\bsm{g}_{1,k}\}^{\rm H} \bsm{S}_{[i]}^{\rm H} \bsm{G}_2^{\rm H} \bsm{\Phi}^{\rm H}  \bsm{\Sigma}^{-1} \bsm{\Phi} \bsm{G}_2 \bsm{S}_{[i]}  \notag\\
&\times  \diag\{\bsm{g}_{1,k}\}\bsm{\theta} +  \bsm{\theta}^{\rm T} \diag \left\{ \bsm{S}_{[i]} \bsm{G}_1 \bsm{Q}(\bsm{H}_0^{\rm H} \bsm{\Phi}^{\rm H} - \bsm{I}) \bsm{\Sigma}^{-1}  \bsm{\Phi}\bsm{G}_2  \right\}  \notag\\
&+  \left(\bsm{\theta}^{\rm T} \diag \left\{ \bsm{S}_{[i]} \bsm{G}_1 \bsm{Q}(\bsm{H}_0^{\rm H} \bsm{\Phi}^{\rm H} - \bsm{I}) \bsm{\Sigma}^{-1}  \bsm{\Phi}\bsm{G}_2 \right\}  \right)^{\rm H},
\end{align}
where the terms irrelevant to $\bsm{\Theta}$ are omitted, and in the derivation we use the facts that $\tr(\bsm{A}\bsm{B})= \tr(\bsm{B}\bsm{A})$,
$\tr\left( \diag\{\bsm{\alpha}\}^{\rm H} \bsm{A} \diag\{\bsm{\alpha}\} \right) = \bsm{\alpha}^{\rm H} [\bsm{A}]_{\diag} \bsm{\alpha}$, and $\tr(\diag\{\bsm{\alpha}\} \bsm{A} ) = \bsm{\alpha}^{\rm T} \diag\{\bsm{A}\}$.
Denote
\begin{align}
&\bsm{\Lambda}_{[i]} = \sum_{k=1}^{K} p_k \diag\{\bsm{g}_{1,k}\}^{\rm H} \bsm{S}_{[i]}^{\rm H} \bsm{G}_2^{\rm H} \bsm{\Phi}^{\rm H} \bsm{\Sigma}^{-1}  \bsm{\Phi} \bsm{G}_2 \bsm{S}_{[i]} \diag\{\bsm{g}_{1,k}\} \\
&\bsm{\alpha}_{[i]} = \left( \diag\left\{ \bsm{S}_{[i]} \bsm{G}_1 \bsm{Q}(\bsm{H}_0^{\rm H} \bsm{\Phi}^{\rm H} - \bsm{I}) \bsm{\Sigma}^{-1}  \bsm{\Phi}\bsm{G}_2  \right\}\right)^{\ast}.
\end{align}
Then, the optimization problem in \eqref{optimal.a2} is converted to
\begin{subequations} \label{optimal.a4}
\begin{align}
 \min_{\bsm{\Theta}} \quad  &\frac{1}{\ell_{\bsm{s}}} \sum_{i=1}^{\ell_{\bsm{s}}} \left( \bsm{\theta}^{\rm H} \bsm{\Lambda}_{[i]} \bsm{\theta} +  \bsm{\alpha}_{[i]}^{\rm H}\bsm{\theta} + \bsm{\theta}^{\rm H} \bsm{\alpha}_{[i]} \right)\label{optimal.a4.a} \\
~\textrm{s.t.} ~\quad  &|\theta_n| = 1, \forall n \in \mathcal{I}_N
\end{align}
\end{subequations}
where \eqref{optimal.a4.a} utilizes the fact that $\bsm{\alpha}_{[i]}^{\rm T}\bsm{\theta}^{\ast} + \bsm{\theta}^{\rm T} \bsm{\alpha}_{[i]}^{\ast} =\bsm{\alpha}_{[i]}^{\rm H}\bsm{\theta} + \bsm{\theta}^{\rm H}\bsm{\alpha}_{[i]}$.

The optimization problem in \eqref{optimal.a4} is a non-convex quadratically constrained quadratic
program (QCQP). Following \cite{yan2019passive}, we reformulate \eqref{optimal.a4} as a homogeneous
QCQP by introducing an auxiliary variable $\ell$, i.e.,
\begin{subequations} \label{optimal.a5}
\begin{align}
& \min_{\bar{\bsm{\theta}}} \quad \bar{\bsm{\theta}}^{\rm H}\bsm{R} \bar{\bsm{\theta}} \\
& ~~\textrm{s.t.} \quad~  |\theta_n| = 1, \forall n \in \mathcal{I}_N
\end{align}
\end{subequations}
where $ \bar{\bsm{\theta}} = \begin{bmatrix} \bsm{\theta} \\ \ell \end{bmatrix}$ and
$\bsm{R} =  \frac{1}{\ell_{\bsm{s}}} \sum_{i=1}^{\ell_{\bsm{s}}} \begin{bmatrix} \bsm{\Lambda}_{[i]} &
\bsm{\alpha}_{[i]} \\
 \bsm{\alpha}_{[i]}^{\rm H} & 0 \end{bmatrix}$.
The optimization of \eqref{optimal.a5} is generally an NP-hard problem \cite{so2007approximating}.
We further simplify problem \eqref{optimal.a5} as a standard semidefinite program (SDP) by noting $\bar{\bsm{\theta}}^{\rm H}\bsm{R} \bar{\bsm{\theta}} = \textrm{tr}[\bsm{R} \bsm{\Delta}]$,
where $ \bsm{\Delta} = \bar{\bsm{\theta}}\bar{\bsm{\theta}}^{\rm H} $. $\bsm{\Delta}$ is a rank-$1$ positive semidefinite matrix, i.e., $\bsm{\Delta}\succcurlyeq 0$ and $\text{rank}(\bsm{\Delta})=1$. By relaxing the rank-one constraint on $\bsm{\Delta}$, we obtain
\begin{subequations} \label{optimal.a6}
\begin{align}
 \min_{\bsm{\Delta}} \quad &\textrm{tr}( \bsm{R}\bsm{\Delta} ) \\
~\textrm{s.t.} \quad~  &\bsm{\Delta} \succcurlyeq 0; \Delta_{n,n}=1, \forall n \in \mathcal{I}_{N+1}.
\end{align}
\end{subequations}
The SDP problem in \eqref{optimal.a6} can be solved by the existing convex optimization solvers such as CVX\cite{grant2014cvx}.

There is no guarantee that the optimal $\bsm{\Delta}$ of the SDP in \eqref{optimal.a6} is rank-one in general.
We obtain a feasiable solution of $\bar{\bsm{\theta}}$ from $\bsm{\Delta}$ by using the Gaussian randomization method \cite{so2007approximating}. Specifically, we first take the eigenvalue decomposition of $\bsm{\Delta}$ as
$\bsm{\Delta} = \bsm{U}\bsm{\Sigma}\bsm{U}^{\rm H}$, where $\bsm{U} \in \mathbb{C}^{(N+1) \times (N+1)}$ is a unitary matrix and $\bsm{\Sigma}\in \mathbb{C}^{(N+1) \times (N+1)}$ is a diagonal matrix.
Then, a feasible solution of $\bar{\bsm{\theta}}$ is given by
$\bar{\bsm{\theta}} = \bsm{U}\bsm{\Sigma}^{1/2}\bsm{r}$, where $\bsm{r} \in \mathbb{C}^{N+1} $ is a random vector
with each element generated from the circularly symmetric complex Gaussian (CSCG) distribution $\mathcal{CN}(0, 1)$.
To find a better $\bar{\bsm{\theta}}$, we generate $\bsm{r}$ by $\ell_{\bsm{\theta}}$  times and choose the one with the minimum objective value of \eqref{optimal.a5}.
Finally, a suboptimal solution of $\bsm{\theta}$ to problem \eqref{optimal.a2} is given by
\begin{align}  \label{optimal.Theta}
\bsm{\theta} = \frac{\left[\bar{\bsm{\theta}}\right]_{(1:N)}/\bar{\theta}_{N+1}}
{\left\|\left[\bar{\bsm{\theta}}\right]_{(1:N)}/\bar{\theta}_{N+1}\right\|_2},
\end{align}
where $[\bsm{a}]_{(1:N)}$ denotes the vector that contains the first $N$ elements of $\bsm{a}$.

\subsection{Solve $\{\mathcal{Q}(\bsm{\Theta},\bsm{s}_{[i]} )\}$ for Given $\bsm{\Theta}$}

We now solve $\mathcal{Q}(\bsm{\Theta},\bsm{s}_{[i]} ) = \min_{\bsm{\Phi},\bsm{\Sigma}} \mathcal{J}_{[i]}(\bsm{\Phi},\bsm{\Sigma}), \forall i \in \mathcal{I}_{\ell_{\bsm{s}}}$ for given $\bsm{\Theta}$.
The derivative of $ \mathcal{J}_{[i]} (\bsm{\Phi},\bsm{\Sigma})$ with respect to $\bsm{\Phi} $ is given by
\begin{align}
\frac{ \partial\mathcal{J}_{[i]} (\bsm{\Phi},\bsm{\Sigma})}{\partial \bsm{\Phi} } &= -2 \bsm{\Sigma}^{-1} \mbs{E}_{\bsm{y},\bsm{x}|\bsm{s}_{[i]}} [ \bsm{x}- \bsm{\Phi}\bsm{y}_{[i]}]\bsm{y}_{[i]}^{\rm H} \notag\\
&= -2 \bsm{\Sigma}^{-1}( \bsm{\mathcal{C}}_{\bsm{x}\bsm{y}|\bsm{s}_{[i]}}- \bsm{\Phi}\bsm{\mathcal{C}}_{\bsm{y}\bsm{y}|\bsm{s}_{[i]}}) ,
\end{align}
where $ \bsm{\mathcal{C}}_{\bsm{x}\bsm{y}|\bsm{s}_{[i]}}$ is the covariance matrix of $\bsm{x}$ and $\bsm{y}$ conditioned on $\bsm{s}$ given by
\begin{align}
&\bsm{\mathcal{C}}_{\bsm{x}\bsm{y}|\bsm{s}_{[i]}} = \bsm{Q} ( \bsm{G}_2 \bsm{\Theta}\bsm{S}_{[i]}\bsm{G}_1 + \bsm{H}_0)^{\rm H},
\end{align}
 and $ \bsm{\mathcal{C}}_{\bsm{y}\bsm{y}|\bsm{s}_{[i]}}$ is the covariance matrix of $\bsm{y}$ conditioned on $\bsm{s}$ given by
\begin{align}
\boldsymbol{\mathcal{C}}_{\bsm{y}\bsm{y}|\bsm{s}_{[i]}} &=  \bsm{G}_2 \bsm{\Theta}\bsm{S}_{[i]}\bsm{G}_1 \bsm{Q} \bsm{G}_1^{\rm H}\bsm{S}_{[i]}^{\rm H}\bsm{\Theta}^{\rm H} \bsm{G}_2^{\rm H} + \bsm{G}_2 \bsm{\Theta}\bsm{S}_{[i]}\bsm{G}_1 \bsm{Q} \bsm{H}_0^{\rm H}  \notag\\
&~~~~~+ \left(\bsm{G}_2 \bsm{\Theta}\bsm{S}_{[i]}\bsm{G}_1 \bsm{Q} \bsm{H}_0^{\rm H} \right)^{\rm H}  +  \bsm{H}_0\bsm{Q}\bsm{H}_0^{\rm H} + \sigma_{w}^2 \bsm{I}.
\end{align}
By setting $\frac{ \partial\mathcal{J}_{[i]}(\bsm{\Phi},\bsm{\Sigma})}{\partial \bsm{\Phi} }=\textbf{0}$,
we obtain the optimal $\bsm{\Phi}$ for given $\bsm{s} = \bsm{s}_{[i]}$ as
\begin{align}  \label{optimal.Phi}
\bsm{\Phi}_{[i]} = \bsm{\mathcal{C}}_{\bsm{x}\bsm{y}|\bsm{s}_{[i]}}
 \bsm{\mathcal{C}}_{\bsm{y}\bsm{y}|\bsm{s}_{[i]}}^{-1}, \forall i \in \mathcal{I}_{\ell_{\bsm{s}}}.
\end{align}
Substituting $ \bsm{\Phi}_{[i]}$ into \eqref{optimal.J} and
letting $\frac{ \partial\mathcal{J}_{[i]}(\bsm{\Sigma}|\bsm{\Phi}_{[i]})}{\partial \bsm{\Sigma} }=\textbf{0}$,
we obtain the optimal $\bsm{\Sigma}$ for given $\bsm{s} = \bsm{s}_{[i]}$ as
\begin{align} \label{optimal.Sigma}
\bsm{\Sigma}_{[i]} &= \bsm{\mathcal{C}}_{\bsm{x}\bsm{x}} - \bsm{\mathcal{C}}_{\bsm{x}\bsm{y}|\bsm{s}_{[i]}} \bsm{\mathcal{C}}_{\bsm{y}\bsm{y}|\bsm{s}_{[i]}}^{-1} \bsm{\mathcal{C}}_{\bsm{y}\bsm{x}|\bsm{s}_{[i]}}, \forall i \in \mathcal{I}_{\ell_{\bsm{s}}}.
\end{align}
Finally, $\{\mathcal{Q}(\bsm{\Theta},\bsm{s}_{[i]} )\}$ can be obtained by
\begin{align} \label{optimal.Q}
&\mathcal{Q}(\bsm{\Theta},\bsm{s}_{[i]} )
= \mathcal{J}_{[i]}(\bsm{\Phi}_{[i]},\bsm{\Sigma}_{[i]}) \notag\\
&= \mbs{E}_{\bsm{y},\bsm{x}|\bsm{s}_{[i]}} \left( \log \det(\bsm{\Sigma}_{[i]} )+ \left\|\left( \bsm{\Sigma}_{[i]}\right)^{-\frac{1}{2}} \left(\bsm{x}- \bsm{\Phi}_{[i]} \bsm{y}_{[i]}\right)\right\|_2^2\right) \notag\\
&~~~~~~~~~~~~~~~~~~~~~~~~~~~~~~~~~~~~~~~~~~~~~~~~~~~~~~~~~~~~~~~~~~~~~~~~~ \forall i \in \mathcal{I}_{\ell_{\bsm{s}}}.
\end{align}

\begin{algorithm}[t]
\caption{ SAA-Based Beamforming Algorithm}
\label{alg:SAA}
\begin{algorithmic}[1]
\STATE Initialize $\bsm{\Theta}$ randomly
\REPEAT
\STATE Generate $\{\bsm{s}_{[1]},\ldots,\bsm{s}_{[\ell_{\bsm{s}}]} \}$ independently based on \eqref{pro.s}
\STATE Compute $\{(\bsm{\Phi}_{[i]},\bsm{\Sigma}_{[i]})\}$ based on \eqref{optimal.Phi} and \eqref{optimal.Sigma}
\STATE Compute $\bsm{\Delta}$ by solving \eqref{optimal.a6}
\STATE Compute $\bar{\bsm{\theta}}$ by using the Gaussian randomization method
\STATE Compute $\bsm{\theta}$ based on \eqref{optimal.Theta}
\UNTIL the objective function in \eqref{optimal.a1.a} is reduced by less than $\epsilon_{bf1}$, or the iteration
index reaches $ iter_{bf1}$
\end{algorithmic}
\end{algorithm}

\subsection{Overall Iterative Algorithm}

The overall iterative algorithm is summarized as the pseudocode in Algorithm~\ref{alg:SAA}. We refer to this algorithm as the SAA-based beamforming algorithm. The convergence of the SAA-based algorithm is guaranteed since $I(\bsm{x};\bsm{y}|\bsm{s}) $ increases monotonically in each update of $\bsm{\Theta}$ and $\{\mathcal{Q}(\bsm{\Theta},\bsm{s}_i )\}$.
However, we observe from numerical experiments that the convergence of this algorithm is relatively slow, which incurs a high computational cost.
This inspire us to develop a simplified beamforming method to strike a balance between complexity and performance, as detailed in the next section.

\section{Simplified Beamforming Method } \label{sec.beamf2}

In this section, we present the simplified beamforming method by slightly modifying the target function of problem \eqref{optimal.e}.
Specifically,
by exchanging the order of the minimization over $(\bsm{\Phi}, \bsm{\Sigma})$ and the expectation over $\bsm{s}$, we can recast the beamforming design problem in \eqref{optimal.e} as
\begin{subequations} \label{optimal.b1}
\begin{align}
 \min_{\bsm{\Phi},\bsm{\Sigma},\bsm{\Theta}} \quad  &\mbs{E}_{\bsm{y},\bsm{x},\bsm{s}} \left( \log \det(\bsm{\Sigma})+ \left\|\bsm{\Sigma}^{-\frac{1}{2}} \left(\bsm{x}- \bsm{\Phi}\bsm{y}\right)\right\|_2^2\right) \label{optimal.b1.a} \\
 \textrm{s.t.} ~\quad~  &|\theta_n| = 1, \forall n \in \mathcal{I}_{N}.
\end{align}
\end{subequations}
It is not difficult to see that the expectation over $\bsm{y},\bsm{x},\bsm{s}$ in \eqref{optimal.b1.a} can be solved explicitly, and so the problem formulated in \eqref{optimal.b1} is computationally more friendly to handle than the one in \eqref{optimal.a1}.
Yet, the problem in \eqref{optimal.b1} is still non-convex, and it is in general difficult to find an exact solution to \eqref{optimal.b1}. Similarly to the approach described in the preceding section, we propose an alternating optimization method that iteratively optimize $\bsm{\Theta}$ and $(\bsm{\Phi},\bsm{\Sigma})$ to find a suboptimal solution of \eqref{optimal.b1}. The details of the algorithm are described below.

\subsection{ Optimal $\bsm{\Theta}$ for Fixed $(\bsm{\Phi},\bsm{\Sigma})$}
\label{sys.theta2}

For fixed $(\bsm{\Phi},\bsm{\Sigma})$, the optimization problem in \eqref{optimal.b1} reduces to
\begin{subequations} \label{optimal.b2}
\begin{align}
 \min_{\bsm{\Theta}} \quad  &\mbs{E}_{\bsm{y},\bsm{x},\bsm{s}} \left\|\bsm{\Sigma}^{-\frac{1}{2}}\left(\bsm{x}- \bsm{\Phi}\bsm{y}\right)\right\|_2^2 \\
 \textrm{s.t.} ~\quad  &|\theta_n| = 1, \forall n \in \mathcal{I}_{N}.
\end{align}
\end{subequations}
Problem \eqref{optimal.b2} can be equivalently rewritten as
\begin{subequations} \label{optimal.b7}
\begin{align}
\min_{\bsm{\Theta}} \quad  &\bsm{\theta}^{\rm H} \bsm{\Lambda} \bsm{\theta} +  \bsm{\alpha}^{\rm H}\bsm{\theta} + \bsm{\theta}^{\rm H} \bsm{\alpha} \\
 \textrm{s.t.} ~\quad  &|\theta_n| = 1, \forall n \in \mathcal{I}_{N}
\end{align}
\end{subequations}
where
\begin{align}
\bsm{\Lambda} &= \sum_{k=1}^{K} \rho^2 p_k \diag\{\bsm{g}_{1,k}\}^{\rm H}\bsm{G}_2^{\rm H} \bsm{\Phi}^{\rm H} \bsm{\Sigma}^{-1}  \bsm{\Phi} \bsm{G}_2 \diag\{\bsm{g}_{1,k}\}  \notag\\
&~~~~+ \rho(1-\rho) \diag\left\{ \bsm{V}^{\rm H} \bsm{G}_2^{\rm H} \bsm{\Phi}^{\rm H} \bsm{\Sigma}^{-1}  \bsm{\Phi} \bsm{G}_2 \bsm{V} \right\} \label{optimal.b21}, \\
&\bsm{\alpha} = \rho\left(\diag\left\{\bsm{G}_1 \bsm{Q}(\bsm{H}_0^{\rm H} \bsm{\Phi}^{\rm H} - \bsm{I}) \bsm{\Sigma}^{-1}  \bsm{\Phi}\bsm{G}_2  \right\}\right)^{\ast}.\label{optimal.b22}
\end{align}
The detailed derivation of the equivalence of \eqref{optimal.b2} and \eqref{optimal.b7} can be found in Appendix B.

Similarly to \eqref{optimal.a4}, the non-convex QCQP problem in \eqref{optimal.b7} can be approximately solved by converting it to an SDP as
\begin{subequations} \label{optimal.b8}
\begin{align}
 \min_{\bsm{\Delta}} \quad &\textrm{tr}( \bar{\bsm{R}}\bsm{\Delta} ) \\
 ~\textrm{s.t.} \quad~  &\bsm{\Delta} \succcurlyeq 0; \Delta_{n,n}=1, \forall n \in \mathcal{I}_{N+1}
\end{align}
\end{subequations}
where
$\bar{\bsm{R}} =  \begin{bmatrix} \bsm{\Lambda} &
\bsm{\alpha} \\
 \bsm{\alpha}^{\rm H} & 0 \end{bmatrix}$.
The method of obtaining $\bsm{\theta}$ from $\bsm{\Delta}$ is basically the same as the one described in Section~\ref{sys.theta1}.
The details are omitted for brevity.

\subsection{ Optimal $(\bsm{\Phi},\bsm{\Sigma})$ for fixed $\bsm{\Theta}$}

For a fixed $\bsm{\Theta}$, the optimization problem in \eqref{optimal.b1} reduces to
\begin{align} \label{optimal.c1}
& \min_{\bsm{\Phi},\bsm{\Sigma}} \quad  \mathcal{J}(\bsm{\Phi},\bsm{\Sigma})
\end{align}
where
\begin{align}
\mathcal{J}(\bsm{\Phi},\bsm{\Sigma}) = \mbs{E}_{\bsm{y},\bsm{x},\bsm{s}} \left( \log \det(\bsm{\Sigma})+ \left\|\bsm{\Sigma}^{-\frac{1}{2}} \left(\bsm{x}- \bsm{\Phi}\bsm{y}\right)\right\|_2^2\right).
\end{align}
Letting $\frac{ \partial\mathcal{J}(\bsm{\Phi},\bsm{\Sigma})}{\partial \bsm{\Phi} }=\textbf{0}$, we obtain
\begin{align}  \label{optimal.Phi2}
\bsm{\Phi}^{opt} = \mbs{E}_{\bsm{s}} ( \bsm{\mathcal{C}}_{\bsm{x}\bsm{y}|\bsm{s}} )
 \left[\mbs{E}_{\bsm{s}}( \bsm{\mathcal{C}}_{\bsm{y}\bsm{y}|\bsm{s}} ) \right]^{-1},
\end{align}
where
\begin{align}
&\mbs{E}_{\bsm{s}} ( \bsm{\mathcal{C}}_{\bsm{x}\bsm{y}|\bsm{s}} ) = \bsm{Q} (\rho\bsm{G}_2 \bsm{\Theta}\bsm{G}_1 + \bsm{H}_0)^{\rm H},
\end{align}

\begin{align}
&\mbs{E}_{\boldsymbol{s}} \left(\boldsymbol{\mathcal{C}}_{\boldsymbol{y}\boldsymbol{y}|\boldsymbol{s}} \right) = \rho(1-\rho) \boldsymbol{G}_2 \boldsymbol{\Theta} \left[\boldsymbol{G}_1 \boldsymbol{Q} \boldsymbol{G}_1^{\rm H}\right]_{\diag} \boldsymbol{\Theta}^{\rm H} \boldsymbol{G}_2^{\rm H} \notag\\
&~~~~~~~~~~~~~~ + \rho^2 \boldsymbol{G}_2 \boldsymbol{\Theta} \boldsymbol{G}_1 \boldsymbol{Q} \boldsymbol{G}_1^{\rm H}\boldsymbol{\Theta}^{\rm H} \boldsymbol{G}_2^{\rm H} + \rho\boldsymbol{G}_2 \boldsymbol{\Theta}\boldsymbol{G}_1 \boldsymbol{Q} \boldsymbol{H}_0^{\rm H}  \notag\\
&~~~~~~~~~~~~~~ + \rho\left(\boldsymbol{G}_2 \boldsymbol{\Theta}\boldsymbol{G}_1 \boldsymbol{Q} \boldsymbol{H}_0^{\rm H}\right)^{\rm H} + \boldsymbol{H}_0\boldsymbol{Q}\boldsymbol{H}_0^{\rm H} + \sigma_{w}^2 \boldsymbol{I}.
\end{align}
Substituting $ \bsm{\Phi}^{opt}$ into \eqref{optimal.c1} and
letting $\frac{ \partial\mathcal{J}(\bsm{\Sigma}|\bsm{\Phi}^{opt})}{\partial \bsm{\Sigma} }=\textbf{0}$,
we obtain
\begin{align} \label{optimal.Sigma2}
&\bsm{\Sigma}^{opt} =  \bsm{\mathcal{C}}_{\bsm{x}\bsm{x}} - \mbs{E}_{\bsm{s}} ( \bsm{\mathcal{C}}_{\bsm{x}\bsm{y}|\bsm{s}} )
 \mbs{E}_{\bsm{s}}( \bsm{\mathcal{C}}_{\bsm{y}\bsm{y}|\bsm{s}} )^{-1} \mbs{E}_{\bsm{s}} (\bsm{\mathcal{C}}_{\bsm{y}\bsm{x}|\bsm{s}} ) .
\end{align}

\subsection{Overall Iterative Algorithm}

The overall iterative algorithm is summarized as the pseudocode in Algorithm~\ref{alg:SBM}. The convergence of Algorithm~\ref{alg:SBM} is guaranteed since the target function of \eqref{optimal.b1} decreases monotonically in each update of $\bsm{\Theta}$ and $(\bsm{\Phi},\bsm{\Sigma})$.
We now briefly compare the computational complexities of the simplified beamforming algorithm (i.e., Algorithm~\ref{alg:SBM}) and the SAA-base beamforming algorithm (i.e., Algorithm~\ref{alg:SAA}). By inspection, the only difference of Algorithm~\ref{alg:SBM} is to replace Lines $3$ and $4$ of Algorithm~\ref{alg:SAA} by Line $3$ of Algorithm~\ref{alg:SBM}. With this replacement, the complexity of the related part is reduced by $\ell_{\bsm{s}}$ times since no sample average is involved in Line $3$ of Algorithm~\ref{alg:SBM}. Furthermore, it can be seen from the numerical results presented later in Section~\ref{sec.Simul} that the convergence speed of the Algorithm~\ref{alg:SBM} is order-of-magnitude faster than that of Algorithm~\ref{alg:SAA}. This is expected since deterministic optimization algorithms usually converge much faster than stochastic optimization algorithms.

\begin{algorithm}[t]
\caption{Simplified Beamforming Algorithm}
\label{alg:SBM}
\begin{algorithmic}[1]
\STATE Initialize $\bsm{\Theta}$ randomly
\REPEAT
\STATE Compute $(\bsm{\Phi},\bsm{\Sigma})$ according to \eqref{optimal.Phi2} and \eqref{optimal.Sigma2}
\STATE Compute $\bsm{\Delta}$ by solving \eqref{optimal.b8}
\STATE Compute $\bar{\bsm{\theta}}$ by Gaussian randomization method
\STATE Compute $\bsm{\theta}$ based on \eqref{optimal.Theta}
\UNTIL the objective function in \eqref{optimal.b1.a} is reduced by less than $\epsilon_{bf2}$, or the iteration
index reaches $iter_{bf2}$
\end{algorithmic}
\end{algorithm}

\section{Receiver Design} \label{sec.receiver}
\subsection{Problem Formulation}

The receiver at the BS aims to reliably recover both the information from the users and the RIS (i.e., $\bsm{X}$ and $\bsm{s}$).
This recovery problem can be formulated by using the maximum {\it a posteriori} principle as
\begin{align}\label{problem.formu}
\left( \hat{\boldsymbol{X}},\hat{\boldsymbol{s}}\right) = \arg\max_{\boldsymbol{X},\boldsymbol{s}}p_{\boldsymbol{X},\boldsymbol{s}|\boldsymbol{Y}}(\boldsymbol{X},\boldsymbol{s}|\boldsymbol{Y}).
\end{align}

Exactly solving \eqref{problem.formu} is difficult by noting the bilinear model of $\bsm{X}$ and $\bsm{s}$ in \eqref{channel.H1}. Message passing is a powerful technique to yield near-optimal low-complexity solutions to sophisticated inference problems.
Particularly, the PBiGAMP algorithm \cite{parker2016parametric} can possibly handle the bilinear inference problem in \eqref{problem.formu}. To see this,
we rewrite the model in \eqref{channel.H1} as
\begin{align} \label{channel.H3}
\bsm{Y} &= (\bsm{G}_2 \bsm{\Theta}\bsm{S}\bsm{G}_1  + \bsm{H}_0 )\bsm{X} + \bsm{W}  \notag\\
 &= \left(\sum_{n=1}^N  s_n \theta_n \bsm{g}_{2,n} \bsm{g}_{1,n}^{\rm T} + \bsm{H}_0 \right) \bsm{X} + \bsm{W} \notag\\
 &= \sum_{n=0}^N s_n \bsm{H}_n \bsm{X} + \bsm{W},
\end{align}
where $\bsm{H}_n\triangleq \theta_n \bsm{g}_{2,n} \bsm{g}_{1,n}^{\rm T} \in \mathbb{C}^{M \times K}$ is a rank-$1$ matrix $\forall n \in \mathcal{I}_{N}$ and $s_0=1$ is a constant. Since \eqref{channel.H3} is special case of the model \cite[eq.~2]{parker2016parametric}, we see that the PBiGAMP algorithm is indeed applicable to problem \eqref{problem.formu}.

However, there are two issues with the PBiGAMP algorithm when it is applied to solve \eqref{problem.formu}. First, the PBiGAMP algorithm requires that the measurement matrixes consist of the i.i.d. Gaussian elements. In \eqref{channel.H3},  the measurement matrices $\{\bsm{H}_1,\ldots,\bsm{H}_N \}$ are rank-$1$, which violates the i.i.d. assumption in PBiGAMP. Second, the computational complexity of PBiGAMP is $\mathcal{O}(MNKT^2)$, which is unaffordable for a large $T$.

\begin{figure}
  \centering
  \includegraphics[width=3 in]{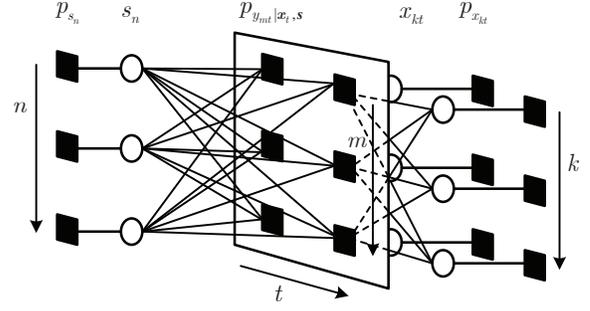}\\
  \caption{The factor graph representation for the joint probability in \eqref{problem.pro} with $M=N=K=3$ and $T=2$.}\label{factor_graph}
\end{figure}

To avoid the above two issues, we develop a turbo message passing algorithm as follows.
We start with the factor graph representation of the probability model involved in \eqref{problem.formu}.
From the Bayes' rule, we obtain
\begin{subequations} \label{problem.pro}
\begin{align}
&p_{\bsm{X},\bsm{s}|\bsm{Y}}(\bsm{X},\bsm{s}|\bsm{Y}) \notag\\
&\varpropto p_{\bsm{Y}|\bsm{X},\bsm{s}}(\bsm{Y}|\bsm{X},\bsm{s})
 p_{\bsm{X}}(\bsm{X}) p_{\bsm{s}}(\bsm{s}) \label{problem.pro.a}\\
&= \left[ \prod_{m=1}^M \prod_{t=1}^T p_{y_{mt}|\bsm{x}_t,\bsm{s} }\left(y_{mt}|\sum_{n=1}^N \sum_{k=1}^K s_n h_{n,mk}x_{k,t} \right)\right] \notag\\
&~~~~\times \left[ \prod_{k=1}^K \prod_{t=1}^T p_{x_{kt}}(x_{kt})\right]\left[\prod_{n=0}^N p_{s_n}(s_n)\right] ,\label{problem.pro.b}
\end{align}
\end{subequations}
where the notation $\propto$ in \eqref{problem.pro.a} means equality up to a constant scaling factor;
\eqref{problem.pro.b} is from the fact that $\{x_{kt}\}$ and $\{s_n\}$ are independent of each other.
The factorized posterior distribution in \eqref{problem.pro.b} can be represented by a factor graph, as depicted in Fig.~\ref{factor_graph}.
Note that in Fig.~\ref{factor_graph}, we use a hollow circle to represent a ``variable node'' and a solid square to represent a ``factor node''.
\begin{figure}
  \centering
  \includegraphics[width=3.5 in]{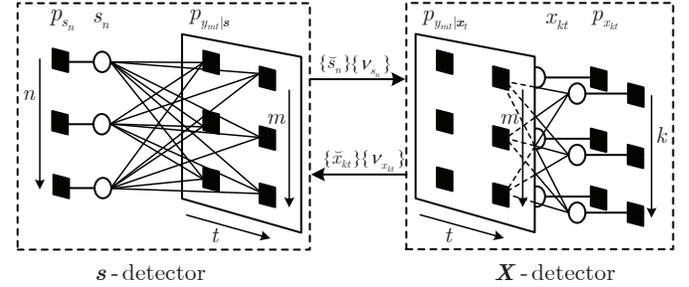}\\
  \caption{The factor graph representation for turbo message passing with $M=N=K=3$ and $T=2$.}\label{turbo_graph}
\end{figure}

We are now ready to present the turbo message passing framework by dividing the whole factor graph in Fig.~\ref{factor_graph} into two modules, namely the $\bsm{X}$-detector and the $\bsm{s}$-detector, as depicted in Fig.~\ref{turbo_graph}.
We iteratively estimate $\bsm{X}$ and $\bsm{s}$ by performing message passing in each module.
The messages are exchanged between the two modules until convergence or the maximum iteration number is
reached. The $\bsm{X}$-detector estimates $\bsm{X}$ based on the observed signal $\bsm{Y}$ and the messages from the $\bsm{s}$-detector. The output of the $\bsm{X}$-detector are the
means and variances of $\{x_{kt}\}$ denoted by $\{\breve{x}_{kt}\}$ and $\{\upsilon_{x_{kt}}\}$. Similarly, the $\bsm{s}$-detector estimates $\bsm{s}$ based on the observed signal $\bsm{Y}$ and the messages from the $\bsm{X}$-detector, and outputs the
means and variances of $\{s_n\}$ denoted by $\{\breve{s}_{n}\}$ and $\{\upsilon_{s_{n}}\}$. In the following subsections, we give a detailed description of the turbo massage passing algorithm.

\subsection{Design of $\bsm{X}$-Detector} \label{sec.x}

We first describe the design of the $\bsm{X}$-detector. From Fig.~\ref{turbo_graph}, the input means and variances of $\bsm{s}$ are respectively given by $\breve{\bsm{s}}$ and $\{\upsilon_{s_{n}}\}$.
Let $\tilde{\bsm{s}}=\bsm{s}-\breve{\bsm{s}}$ be the residual interference of $\bsm{s}$ with mean zero and variances $\{\upsilon_{s_{n}}\}$ . Then the observed signal $\bsm{Y}$ can be expressed as
\begin{align} \label{model.x}
&\bsm{Y} =  \sum_{n=0}^N (\breve{s}_n + \tilde{s}_n) \bsm{H}_n \bsm{X} + \bsm{W} = \breve{\bsm{H}} \bsm{X} + \tilde{\bsm{W}}_{\bsm{X}},
\end{align}
where $\breve{\bsm{H}}= \sum_{n=0}^N \breve{s}_n \bsm{H}_n $ is the linear transform matrix known by the $\bsm{X}$-detector and $ \tilde{\bsm{W}}_{\bsm{X}} \triangleq \sum_{n=0}^N \tilde{s}_n \bsm{H}_n \bsm{X} + \bsm{W}$ is the equivalent noise.
Assume that the elements of $\tilde{\bsm{W}}_{\bsm{X}}$ are independent of each other and obey a CSCG distribution. Denote by $\tilde{w}_{\bsm{X},mt}$ the $(m,t)$th element of $\tilde{\bsm{W}}_{\bsm{X}}$.
The variance of each $\tilde{w}_{\bsm{X},mt}$ is
\begin{align} \label{variance.x}
&\upsilon_{\tilde{w}_{\bsm{X},mt}}= \sum_{i=0}^N \sum_{k=1}^K \upsilon_{s_i}  |h_{i,mk}|^2 p_k + \sigma_w^2, \forall m,t,
\end{align}
where $p_k$ is the power of user $k$.
The derivation of \eqref{variance.x} is given in Appendix C-A.
Thus, we have
\begin{align}
&p_{y_{mt}|\bsm{x}_t}(y_{mt}|\bsm{x}_t) \notag\\
&= \mathcal{CN}\left(y_{mt};\sum_{n=1}^N \sum_{k=1}^K \breve{s}_n h_{n,mk}x_{k,t}, \upsilon_{\tilde{w}_{\bsm{X},mt}}\right),\forall m,t.
\end{align}

Given the model in \eqref{model.x}, the recovery of $\bsm{X}$ from the observation $\bsm{Y}$ is a linear estimation problem.
The existing algorithm GGAMP-SBL \cite{al2017gamp}
is designed to handle such a linear estimation problem with an arbitrary measurement matrix, and thus can be used here for the recovery of $\bsm{X}$.
\begin{algorithm}[!htbp]
\vspace{0.2mm}
\caption{\textbf{\!:} \vspace{0.15mm} The GGAMP-SBL algorithm}
\label{alg.GGAMP-SBL}
\begin{algorithmic}[1]
\REQUIRE $\bsm{Y}$, $\breve{\bsm{H}}$, $\{\upsilon_{\tilde{w}_{\bsm{X},mt}} \}$, and $\mathcal{CN}(\cdot; \breve{x}_{kt},\upsilon_{x_{kt}}), \forall k,t$.  \\
\hspace{-0.6cm}Initialization: $\forall m,k,t :$ $\breve{x}_{kt}^{(0)}=\breve{x}_{kt}$, $\upsilon_{x_{kt}}^{(0)}=\upsilon_{x_{kt}}$,
  $\breve{\nu}_{x_{kt}}^{(0)}=\upsilon_{x_{kt}}$, $ \breve{u}_{mt}^{(0)}=0$.\\
\hspace{-0.6cm}for $i = 1,2,\ldots, I_{max}^{x}$ \quad \% EM iteration \\
Initialization: $\forall m,k,t :$ $\hat{x}_{k,t}^{(0)} = \breve{x}_{kt}^{(i-1)}$, $\nu_{x_{k,t}}^{(0)} = \breve{\nu}_{x_{kt}}^{(i-1)}$,
$ \hat{u}_{mt}^0 = \breve{u}_{mt}^{(i-1)}$
\quad \%E-Step\\
for $\ell = 1,2,\ldots, \ell_{max}^{x}$   \quad \% GGAMP iteration  \\
\STATE\hspace{0.5cm}$\forall m, t:$ $\nu_{p_{mt}}^{(\ell)} = \sum_{k=1}^k |\breve{h}_{mk}|^2 \nu_{x_{k,t}}^{(\ell-1)}$\\
\STATE\hspace{0.5cm}$\forall m, t:$ $\hat{p}_{mt}^{(\ell)} = \sum_{k=1}^k \breve{h}_{mk} \hat{x}_{(kt)}^{(\ell-1)} - \nu_{p_{mt}}^{(\ell)} \hat{u}_{mt}^{(\ell-1)} $ \\ 
\STATE\hspace{0.5cm}$\forall m, t:$ $\nu_{z_{mt}}^{(\ell)} = \mbs{Var} (z_{mt} | y_{mt}; \hat{p}_m^{(\ell)}, \nu_{p_{mt}}^{(\ell)})$\\ \STATE\hspace{0.5cm}$\forall m, t:$ $\hat{z}_{mt}^{(\ell)} = \mbs{E}(z_{mt} | y_{mt}; \hat{p}_{mt}^{(\ell)}, \nu_{p_{mt}}^{(\ell)})$ \\ 
\STATE\hspace{0.5cm}$\forall m, t:$ $\nu_{u_{mt}}^{(\ell)} = \left(1-\nu_{z_{mt}}^{(\ell)} / \nu_{p_{mt}}^{(\ell)} \right) / \nu_{p_{mt}}^{(\ell)} $\\
\STATE\hspace{0.5cm}$\forall m, t:$ $\hat{u}_{mt}^{(\ell)} =  (1-\xi_u^x)\hat{u}_{mt}^{(\ell-1)} + \xi_u^x \left( \hat{z}_{mt}^{(\ell)} - \hat{p}_{mt}^{(\ell)} \right) \nu_{p_{mt}}^{(\ell)}$ \\
\STATE\hspace{0.5cm}$\forall k, t:$ $\nu_{r_{kt}}^{(\ell)} = \left( \sum_{m=1}^M |\breve{h}_{mk}|^2 \nu_{u_{mt}}^{(\ell)} \right)^{-1} $\\
\STATE\hspace{0.5cm}$\forall k, t:$ $\hat{r}_{kt}^{(\ell)} = \hat{x}_{kt}^{(\ell-1)} + \nu_{r_{kt}}^{(\ell)} \sum_{m=1}^M \breve{h}_{mk} \hat{u}_{mt}^{(\ell)} $ \\
\STATE\hspace{0.5cm}$\forall k, t:$ $\nu_{x_{kt}}^{(\ell)} = \mbs{Var} (x_{kt} | \bsm{y}_t; \hat{r}_{kt}^{(\ell)}, \nu_{r_{kt}}^{(\ell)} )$\\
\STATE\hspace{0.5cm}$\forall k, t:$ $\hat{x}_{kt}^{(\ell)} = (1-\xi_x^x)\hat{x}_{kt}^{(\ell-1)} + \xi_x^x \mbs{E}(x_{kt} | \bsm{Y}; \hat{r}_{kt}^{(\ell)}, \nu_{r_{kt}}^{(\ell)})$ \\ \vspace{0.6mm}
\STATE\hspace{0.4cm} if $\|\hat{x}_{kt}^{(\ell)} - \hat{x}_{kt}^{(\ell-1)}\|^2/\|\hat{x}_{kt}^{(\ell)}\|^2 \leq \epsilon_{\text{gamp}}^x$, break \\
end for\\
\STATE$\forall m,k,t :$ $\breve{x}_{kt}^{(i)}= \hat{x}_{kt}^{(\ell)}$, $\breve{\nu}_{x_{k,t}}^{(i)}= \nu_{x_{kt}}^{(\ell)}$,
 $ \breve{u}_{mt}^{(i)}=\hat{u}_{mt}^{(\ell)}$  \\
\STATE$\forall k,t :$ $\upsilon_{x_{kt}}^{(i)} = |\breve{x}_{kt}^{(i)}|^2 + \breve{\nu}_{x_{k,t}}^{(i)} $ \quad \%M-Step\\
\STATE if $\|\hat{x}_{kt}^{(i)} - \hat{x}_{kt}^{(i-1)}\|^2/\|\hat{x}_{kt}^{(i)}\|^2 \leq \epsilon_{\text{em}}^x$, break \\
\hspace{-0.6cm}end for\\
\ENSURE : $\{\breve{x}_{kt}^{(i)}\}$ and $\{\breve{\nu}_{x_{kt}}^{(i)}\}$.
\end{algorithmic}
\end{algorithm}


For completeness, the details of the GGAMP-SBL algorithm are included as Algorithm~\ref{alg.GGAMP-SBL}. The algorithm
 uses a Gaussian distribution $\mathcal{CN}(\cdot; \breve{x}_{kt},\upsilon_{x_{kt}}),\forall k,t$ to approximate the prior distribution of each $x_{kt}$ with mean $\breve{x}_{kt}$ and variance $\upsilon_{x_{kt}}$. The variances $\{\upsilon_{x_{kt}}\}$ are updated in the M-step of each expectation maximization (EM) iteration by step $8$. The E-step estimates the marginal posterior distributions of $\{x_{kt}\}$ by the GGAMP algorithm, as given in steps $1$-$11$. In specific,
step $1$ updates the estimates of $\{z_{mt}\}$ with the variances $\{\nu_{p_{mt}}^{(l)}\}$ and means $\{\hat{p}_{mt}^{(l)}\}$ by accumulating the
messages from variable nodes $\{x_{kt}\}$ to check nodes $\{p_{y_{mt}}\}$.
 Step $2$ gives the estimates of the marginal posterior variances $\{\nu_{z_{nt}}^{(l)}\}$ and means $\{\hat{z}_{mt}^{(l)}\}$ of $\{z_{mt}\}$.
 Step $3$ calculates the inverse-residual-variances $\{\nu_{u_{n,t}}^{(l)}\}$ and the scaled residuals $\{\hat{u}_{nt}^{(l)}\}$.
Step $4$ updates the estimates of $\{x_{kt}\}$ with variances $\{\nu_{r_{kt}}^{(l)}\}$ and the means $\{\hat{r}_{kt}^{(l)}\}$ by accumulating the
messages from check nodes $\{p_{y_{mt}}\}$  to variable nodes $\{x_{kt}\}$.
 Step $5$ updates the marginal posterior variances $\{\nu_{x_{kt}}^{(l)}\}$ and means $\{\hat{x}_{kt}^{(l)}\}$ of $\{x_{kt}\}$.
Steps $6$ and $9$ define the termination conditions of the GGAMP iteration and the EM iteration, respectively, where $\epsilon_{\text{gamp}}$ and
$\epsilon_{\text{em}}$ are the corresponding tolerance parameters. $\ell_{max}^x$ and $I_{max}^x$ are the maximum numbers of the GGAMP iteration and the EM iteration, respectively.
The outputs of the GGAMP-SBL algorithm are the marginal posterior means $\{\breve{x}_{kt}^{(i)}\}$ and variances $\{\breve{\nu}_{x_{kt}}^{(i)}\}$ of the last iteration.

 The damping strategy is used in the steps $6$ and $10$ to ensure the convergence of the GAMP algorithm, where $\xi_u^x \in [0,1]$ and $\xi_x^x \in [0,1]$ are the damping factors of $\hat{u}_{mt}$ and $\hat{x}_{kt}$, respectively.
 The variance and mean in steps $3$ and $4$ are take over the probability distribution $p(z_{mt}|y_{mt}; \hat{p}_m, \nu_{p_{mt}})=\frac{1}{C} p(y_{mt}|\bsm{x}_t) \mathcal{CN}(z_{mt}|\hat{p}_m, \nu_{p_{mt}}) $, where $C$ is a normalization factor.
The variance and mean in steps $9$ and $10$ are take over $p(x_{kt}; \hat{r}_{kt}, \nu_{r_{kt}})=\frac{1}{C} \mathcal{CN}(\cdot; \breve{x}_{kt},\upsilon_{x_{kt}}) \mathcal{CN}(x_{kt}|\hat{r}_{kt}, \nu_{r_{kt}}) $.

\subsection{Design of $\bsm{s}$-Detector}

We now describe the design of the $\bsm{s}$-detector. Recall from Fig.~\ref{turbo_graph} that the input mean and variances of $\bsm{X}$ are respectively given by $\breve{\bsm{X}}$ and $\{\upsilon_{x_{kt}}\}$.
Let $\tilde{\bsm{X}}=\bsm{X}-\breve{\bsm{X}}$ be the residual interference of $\bsm{X}$ with mean zero and variances $\{\upsilon_{x_{kt}}\}$ . Then the observed signal $\bsm{Y}$ can be expressed as
\begin{align}
& \bsm{Y} =  \sum_{n=0}^N s_n \bsm{H}_n (\breve{\bsm{X}} + \tilde{\bsm{X}}) + \bsm{W} = \sum_{n=0}^N s_n \bsm{H}_n \breve{\bsm{X}} +  \tilde{\bsm{W}}_{\bsm{s}},
\end{align}
where $ \tilde{\bsm{W}}_{\bsm{s}} \triangleq \sum_{n=0}^N s_n \bsm{H}_n \tilde{\bsm{X}} + \bsm{W}$ is the equivalent noise for the $\bsm{s}$-detector.
Denoted $\breve{\bsm{A}}_n = \bsm{H}_n\breve{\bsm{X}},\forall n\in \{0,1,\ldots,N\}$ and $\breve{\bsm{A}} = [\text{vec}(\breve{\bsm{A}}_0),\ldots, \text{vec}(\breve{\bsm{A}}_N) ]\in \mathbb{C}^{MT \times (N+1)}$. Then, we have
\begin{align} \label{model.s}
&\text{vec}(\bsm{Y})= \breve{\bsm{A}}\bsm{s} + \text{vec}(\tilde{\bsm{W}}_{\bsm{s} }).
\end{align}
Similarly, assume that the elements of $\tilde{\bsm{W}}_{\bsm{s}}$ are independent of each other and obey a CSCG distribution.
Denote by $\tilde{w}_{\bsm{s},mt}$ the $(m,t)$th element of $\tilde{\bsm{W}}_{\bsm{s}}$.
The variance of each $\tilde{w}_{\bsm{s},mt}$ is given by
\begin{align} \label{variance.s}
\upsilon_{\tilde{w}_{\bsm{s},mt}}&=\rho^2 \sum_{k=1}^K \upsilon_{x_{kt}} \left| \sum_{n=1}^N h_{n,mk} \right|^2 + \sigma_w^2 \notag\\
&~~~~+ \rho(1-\rho)\sum_{k'=1}^K \sum_{n'=1}^N \upsilon_{x_{k't}} \left|h_{n',mk} \right|^2 , \forall m,t.
\end{align}
The derivation of \eqref{variance.s} is given in Appendix C-B.
Thus, we have
\begin{align}
&p_{y_{mt}|\bsm{s}}(y_{mt}|\bsm{s}) \notag\\
&= \mathcal{CN}\left(y_{mt};\sum_{n=1}^N \sum_{k=1}^K s_n h_{n,mk}\breve{x}_{k,t}, \upsilon_{\tilde{w}_{\bsm{s},mt}}\right),\forall m,t.
\end{align}

Given the model in \eqref{model.s}, the recovery of $\bsm{s}$ from $\bsm{Y}$ can also be solved by using the GGAMP-SBL algorithm.
The GGAMP-SBL algorithm described in Section~\ref{sec.x} can be directly used to detect $\bsm{s}$ after changing the
measurement matrix form $\breve{\bsm{H}}$ to $\breve{\bsm{A}}$, the likelihood function from $ p_{\bsm{Y}|\bsm{X}}(\bsm{Y}|\bsm{X})$ to $ p_{\bsm{Y}|\bsm{s}}(\bsm{Y}|\bsm{s})$,
 and the {\it priori} from $\mathcal{CN}(\cdot; \breve{x}_{kt},\upsilon_{x_{kt}}), \forall k,t$ to $\mathcal{CN}(\cdot; \breve{s}_{n},\upsilon_{s_{n}}), \forall n$. Moreover, the parameters $I_{max}^{x}$, $\ell_{max}^{x}$, $\xi_u^x $, $\xi_x^x $,
 $ \epsilon_{\text{gamp}}^x$ and $\epsilon_{\text{em}}^x$ in the $\bsm{X}$-detector are correspondingly  replaced by $I_{max}^{s}$, $\ell_{max}^{s}$, $\xi_u^s $, $\xi_x^s $,  $ \epsilon_{\text{gamp}}^s$ and $\epsilon_{\text{em}}^s$.

\subsection{Algorithm Summary}

\begin{algorithm}[!htbp]
\vspace{0.2mm}
\caption{\textbf{\!:} \vspace{0.15mm} Turbo Message Passing Algorithm}
\label{alg.Turbo-Detection}
\begin{algorithmic}[1]
\REQUIRE $\bsm{Y}$, prior distribution $p_{\bsm{s}}(\bsm{s})$, $p_{\bsm{X}}(\bsm{X})$, and
 $\sigma_w^2$ \\
\hspace{-0.6cm}Initialization: $\forall m,n,k,t:$ $\breve{s}_{n}^{(0)} = \int_{s_{n}} s_{n}p_{s_{n}}(s_{n})$, $\breve{x}_{kt}^{(0)} = \int_{x_{kt}} x_{kt}p_{x_{kt}}(x_{kt})$, $\upsilon_{s_{n}}^{(0)} = \int_{s_{n}} \left|s_{n}-\breve{s}_{n}^{(0)}\right|^2 p_{s_{n}}(s_{n})$,
$\upsilon_{x_{kt}}^{(0)} = \int_{x_{kt}} \left|x_{kt}-\breve{x}_{kt}^{(0)}\right|^2 p_{x_{kt}}(x_{kt})$ \\
\hspace{-0.6cm}for $\tau = 1,2,\ldots, \tau_{max}$ \quad \% Turbo iteration \\

\STATE $\breve{\bsm{H}}^{(\tau)}= \sum_{n=0}^N \breve{s}_{n}^{(\tau-1)} \bsm{H}_n $ \% $\bsm{X}$-detector  \\\vspace{0.4mm}
\STATE $\forall m, t:$ $\nu_{\tilde{w}_{\bsm{X},mt}}^{(\tau)} = \sum_{n=0}^N \sum_{k=1}^K \upsilon_{s_{n}}^{(\tau-1)}  |h_{n,mk}|^2 p_k + \sigma_w^2$ \\
\STATE $\forall k, t:$ $\breve{x}_{kt} =  \breve{x}_{kt}^{(\tau-1)}$, $ \upsilon_{x_{kt}} = \upsilon_{x_{kt}}^{(\tau-1)}$ \\
\STATE  Perform  GGAMP-SBL to estimate $\bsm{X}$ by invoking Algorithm~\ref{alg.GGAMP-SBL}, and output $\{\breve{x}_{kt}^{(\tau)}\}$ and $\{\upsilon_{x_{kt}}^{(\tau)}\}$ \\
\STATE  $\forall k, t:$ $\breve{x}_{kt}^{(\tau)} = \arg \min_{c \in \mathcal{C}} | c - \breve{x}_{kt}^{(\tau)} |^2$

\STATE $\forall n:$ $\breve{\bsm{A}}_n^{(\tau)} = \bsm{H}_n \breve{\bsm{X}}^{(\tau)}$,\\
 $\breve{\bsm{A}}^{(\tau)} = [\text{vec}(\hat{\bsm{A}}_0)^{(\tau)},\ldots \text{vec}(\hat{\bsm{A}}_N)^{(\tau)} ]$. \% $\bsm{s}$-detector\\
\STATE $\forall m, t:$ $\nu_{\tilde{w}_{\bsm{s},mt}}^{(\tau)} = \rho^2 \sum_{k=1}^K \upsilon_{x_{kt}}^{(\tau)} \left| \sum_{n=1}^N h_{n,mk} \right|^2 + \rho(1-\rho)\sum_{k'=1}^K \sum_{n'=1}^N \upsilon_{x_{k't}}^{(\tau)} \left|h_{n',mk} \right|^2 + \sigma_w^2$ \\
\STATE $\forall n:$ $\breve{s}_{n} = \breve{s}_{n}^{(\tau-1)}$, $ \upsilon_{s_{n}} = \upsilon_{s_{n}}^{(\tau-1)}$ \\
\STATE Perform GGAMP-SBL to estimate $\bsm{s}$, and output $\{\breve{s}_{n}^{(\tau)}\}$ and $\{\upsilon_{s_{n}}^{(\tau)}\}$ \\
\STATE  $\forall n:$ $\breve{s}_{n}^{(\tau)} = \arg \min_{c \in \{0,1\}} | c - \breve{s}_{n}^{(\tau)} |^2$
\STATE if $\left| \left\| \bsm{Y} - \sum_{n=0}^N \breve{s}_{n}^{(\tau)} \bsm{H}_n \breve{\bsm{X}}^{(\tau)} \right\|_{\rm F}^2/M/T - \sigma_w^2 \right|^2 \leq \epsilon_{\text{td}}$, break \\
\hspace{-0.6cm}end for\\
\ENSURE : $\{\breve{x}_{kt}^{(\tau)}\}$ and $\{\breve{s}_{n}^{(\tau)}\}$
\end{algorithmic}
\end{algorithm}

The turbo message passing algorithm is summarized in Algorithm~\ref{alg.Turbo-Detection}.
In specific, step $1$ updates $ \breve{\bsm{H}} $. Step $2$ updates the variances $\{ \nu_{\tilde{w}_{\bsm{X},mt}}\}$.
Step $3$ initializes the means $ \{\breve{x}_{kt}\}$ and variances $\{\upsilon_{x_{kt}}\}$,
Step $4$ perform the GGAMP-SBL algorithm to update the estimate of $\bsm{X}$. Step $5$ maps each $\breve{x}_{kt}^{(\tau)}$  to the nearest constellation point.
Step $6$ updates $ \breve{\bsm{A}}$. Step $7$ updates the variances $\{ \nu_{\tilde{w}_{\bsm{s},mt}}\}$.
Step $8$ initializes the means $ \{\breve{s}_{n}\}$ and variances $\{\upsilon_{s_{n}}\}$,
Step $9$ performs the GGAMP-SBL algorithm to update the estimate of $\bsm{s}$. Step $10$ makes a hard decision on each $\breve{s}_{n}^{(\tau)}$.  Step $11$ defines the termination condition of the turbo iteration, where $\epsilon_{\text{td}}$ is the tolerance parameter, and $\tau_{max}$ is the maximum number of the turbo iteration.
The outputs of the turbo message passing algorithm are the marginal posterior means $\{\breve{x}_{kt}^{(\tau)}\}$ and $\{\breve{s}_{n}^{(\tau)}\}$ of the last iteration.

We now analyse the computational complexity of the turbo message passing algorithm.
The complexity in step $1$ is $\mathcal{O}(MNK)$. The complexity in step $2$ is $\mathcal{O}(MNKT)$.
In the $\bsm{X}$-detector, the complexity of the GGAMP-SBL algorithm is $\mathcal{O}(MKT)$.
The complexity in step $6$ is $\mathcal{O}(MKT)$. The complexity in step $8$ is $\mathcal{O}(MNKT)$.
In the $\bsm{s}$-detector, the complexity of the GGAMP-SBL algorithm is $\mathcal{O}(MNT)$.
Thus, the computational complexity of the turbo message passing algorithm is dominated by the steps $2$ and $8$, and is given by $\mathcal{O}(MNKT)$.


\section{Extension on Multi-RIS } \label{sec.multi}
\begin{figure}[t]
  \centering
  \includegraphics[width=2.6 in]{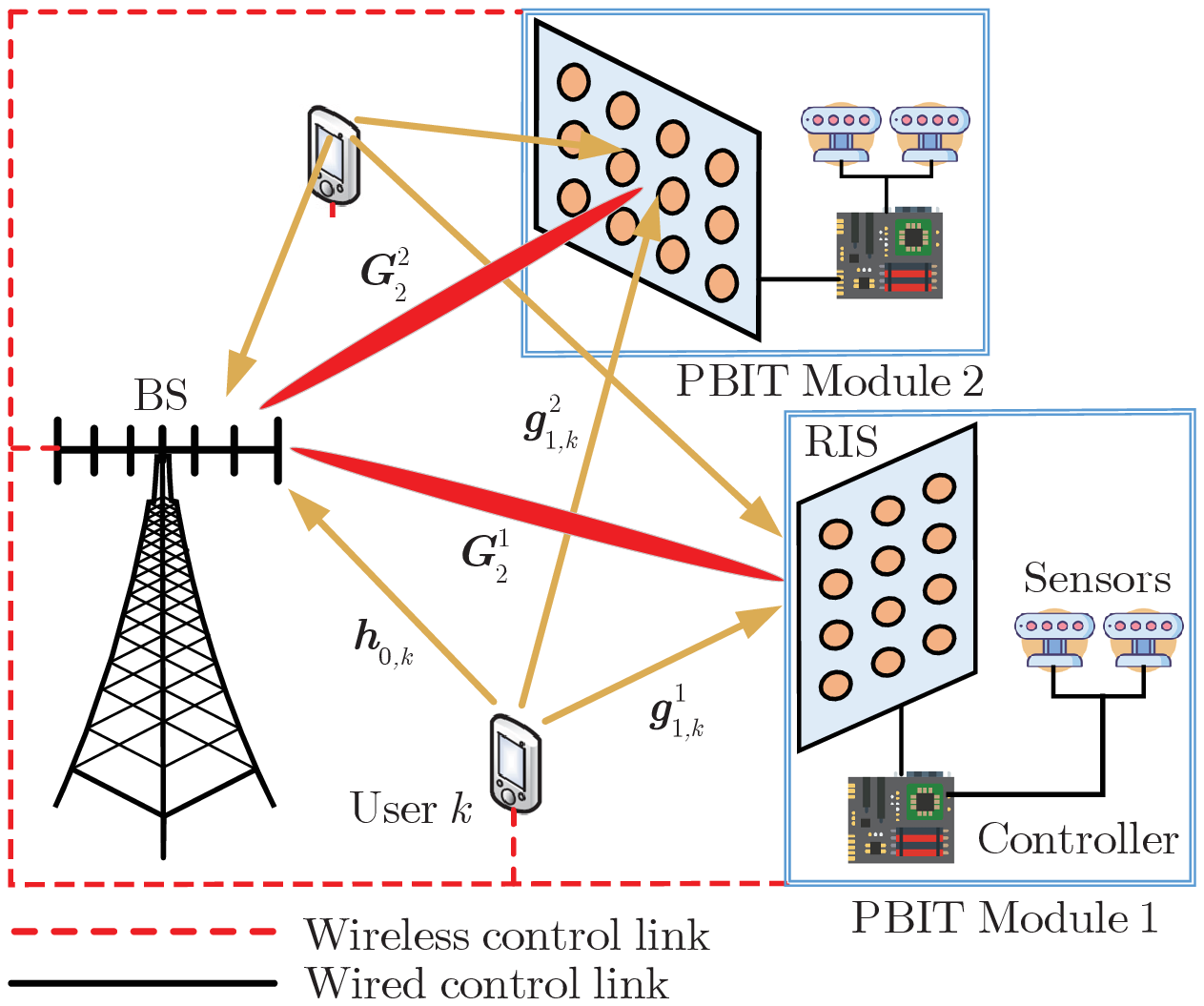}\\
  \caption{A PBIT-enhanced uplink MIMO wireless system based on multi-PBIT.}\label{M-RIS}
\end{figure}

We now extend the discussions from the single-RIS aided multiuser MIMO system to the multi-RIS case.
As illustrated in Fig.~\ref{M-RIS}, we consider an $L$-RIS aided wireless system, where each RIS belongs to a PBIT module equipped with an individual  controller and connected with individual sensors.
The phase shifts of the reflecting elements in all the RISs are jointly designed to enhance the user-BS communication.

The channel model of the multi-RIS aided multiuser MIMO system is described as follows.
The channel of the user-BS link is still denote by $\bsm{H}_0$, where $\bsm{H}_0\in \mathbb{C}^{M \times K}$ is defined in the same way as in the single-RIS system in \eqref{channel.H2}.
Denote by $\text{RIS}^{l}$ the RIS in the $l$th PBIT module. The number of the reflecting elements in $\text{RIS}^{l}$ is $N_l,\forall l\in \mathcal{I}_{L}$.
 Denote by $\bsm{G}_{1}^{l}=[\bsm{g}_{1,1}^l,\ldots,\bsm{g}_{1,N_l}^l]^{\rm T }\in \mathbb{C}^{N_l \times K}$
 the baseband equivalent channel of the user-$\text{RIS}^{l}$ link, where
  $\bsm{g}_{1,n}^l\in \mathbb{C}^{K \times 1} $ represents the channel coefficient vector between all the users and the $n$th reflecting element of $\text{RIS}^{l}$.
   $\bsm{S}^l = \text{diag}\{ \bsm{s}^l\}$ is the diagonal state matrix of $\text{RIS}^{l}$, where
 $\bsm{s}^l = [s_1^l, \ldots, s_{N_l}^l]^{\rm T}$ carries the information of the sensors in the $l$th PBIT module.
 $\bsm{\Theta}^l = \text{diag}\{ \bsm{\theta}^l\}$ is the diagonal phase-shift matrix of $\text{RIS}^{l}$, with
 $ \bsm{\theta}^l = [\theta_1^l, \ldots, \theta_{N_l}^l ]^{\rm T}$.
  Denote by $\bsm{G}_{2}^l=[\bsm{g}_{2,1}^l,\ldots,\bsm{g}_{2,N_l}^l]\in \mathbb{C}^{M \times N_l}$ the baseband equivalent channel of the $\text{RIS}^{l}$-BS link, where  $\bsm{g}_{2,n}^l\in \mathbb{C}^{M \times 1} $ is the channel coefficient vector between the $n$th reflecting element of $\text{RIS}^{l}$ and the BS.
Then, the observed signal matrix in the multi-RIS system over a transmission block can be expressed as
\begin{align} \label{channel.H4}
&\bsm{Y} = \left( \sum_{l=1}^L \bsm{G}_2^l \bsm{\Theta}^l \bsm{S}^l \bsm{G}_1^l  + \bsm{H}_0 \right)\bsm{X} +
 \bsm{W},
\end{align}
where $\bsm{X}$ and $\bsm{W}$ are defined in the same way as in the single-RIS system.

Denoted by $N = \sum_{l=1}^L N_l$ the total number of the reflecting elements in all the $L$ RISs.
$\bsm{G}_1 \triangleq \left[\left(\bsm{G}_1^1\right)^{\rm T},\ldots,\left(\bsm{G}_1^L\right)^{\rm T}\right]^{\rm T} \in \mathbb{C}^{N \times K}$ represents the baseband equivalent channel of the user-RIS link.
$\bsm{S} = \text{diag}\{ \bsm{s}\}$, with $\bsm{s} = \left[\left(\bsm{s}^1\right)^{\rm T},\ldots,\left(\bsm{s}^L\right)^{\rm T}\right]^{\rm T} \in \mathbb{C}^{N \times 1}$, represents the diagonal state matrix of all the RISs.
$\bsm{\Theta} = \text{diag}\{ \bsm{\theta}\}$, where $\bsm{\theta} = \left[\left(\bsm{\theta}^1\right)^{\rm T},\ldots,\left(\bsm{\theta}^L\right)^{\rm T}\right]^{\rm T} \in \mathbb{C}^{N \times 1}$, represents the diagonal phase-shift matrix of all the RISs.
$\bsm{G}_2 \triangleq [\bsm{G}_2^1,\ldots,\bsm{G}_2^L]\in \mathbb{C}^{M \times N}$ represents the baseband equivalent channel of the RIS-BS link.
Then, the system model in the multi-RIS system can still be expressed by \eqref{channel.H1} (which is the same as the single-RIS system).

  Denote by $\rho_l$ the probability that $s^l_n, \forall n \in \mathcal{I}_{N_l}, \forall l \in \mathcal{I}_{L}$ takes the value of $1$, where $s^l_n$ is the state of the $n$th passive element in $\text{RIS}^{l}$.
 Then, the probability of $\bsm{s}$ in the multi-RIS system can be written as
\begin{align}\label{pro.s2}
  p(\bsm{s}) &= \prod_{l=1}^L \left(\prod_{n=1}^{N_l} p(s_n)\right) \notag\\
  &= \prod_{l=1}^L \left(\prod_{n=1}^{N_l} \left( (1-\rho_l)\delta(s_n^l) + \rho_l \delta(s_n^l-1) \right)\right).
\end{align}
Then, we have
\begin{align}
& \mbs{E}_{\bsm{s}}\left(\bsm{S}\right) = \diag\{\mathcal{E}_s\} \\
&\mbs{E}_{\bsm{s}}\left( \bsm{S} \bsm{X}\bsm{S}^{\rm H} \right) = (\mathcal{E}_s \mathcal{E}_s^{\rm H})\odot \bsm{X} \notag\\
&~~~~~~~~~~~~~~~~~~~~~~~~~~~~+ \diag\{\mathcal{E}_s(\textbf{1}_{N}-\mathcal{E}_s)\} \left[\bsm{X} \right]_{\diag},
\label{optimal.2b5}
\end{align}
where $ \mathcal{E}_s \triangleq [\rho_1 \textbf{1}_{N_1}^{\rm T},\ldots, \rho_L \textbf{1}_{N_L}^{\rm T}]^{\rm T} $.

We now discuss the passive beamforming design and the receiver design for the multi-RIS aided multiuser MIMO system.
Note that the system model of the multi-RIS system shares the same expression (i.e., \eqref{channel.H1}) as that of the single-RIS system.
 The only difference is the PDF of $\bsm{s}$ in \eqref{pro.s2} for the multi-RIS case as compared to \eqref{pro.s} for the single-RIS case. This difference induces the following modifications when extending the beamforming and receiver design of the single-RIS case to the multi-RIS case:
\begin{itemize}
  \item {{\it SAA-based beamforming algorithm:}} The independently generated samples $\{\bsm{s}_{[1]},\ldots,\bsm{s}_{[\ell_{\bsm{s}}]} \}$
   are based on the PDF in \eqref{pro.s2} rather than in \eqref{pro.s}.
  \item {{\it Simplified beamforming algorithm:}} We  replace \eqref{optimal.b21} and \eqref{optimal.b22} respectively by
\begin{align}
&\bsm{\Lambda} = \sum_{k=1}^{K} p_k \diag\{\mathcal{E}_s\odot\bsm{g}_{1,k}\}^{\rm H}\bsm{G}_2^{\rm H} \bsm{\Phi}^{\rm H} \bsm{\Sigma}^{-1}  \bsm{\Phi} \bsm{G}_2  \notag\\
&~~~~~~~~\times \diag\{\mathcal{E}_s\odot\bsm{g}_{1,k}\} + \diag\left\{ \bsm{G}_2^{\rm H} \bsm{\Phi}^{\rm H} \bsm{\Sigma}^{-1}  \bsm{\Phi} \bsm{G}_2  \right. \notag\\
&~~~~~~~~ \left. \diag\{\mathcal{E}_s(\textbf{1}_{N}-\mathcal{E}_s)\} \left[\bsm{G}_1 \bsm{Q} \bsm{G}_1^{\rm H} \right]_{\diag} \right\} \label{optimal.lamta2} \\
&\bsm{\alpha} = \rho\left(\diag\left\{\bsm{G}_1 \bsm{Q}(\bsm{H}_0^{\rm H} \bsm{\Phi}^{\rm H} - \bsm{I}) \bsm{\Sigma}^{-1}  \bsm{\Phi}\bsm{G}_2  \right\}\right)^{\ast}.
\end{align}
  \item {{\it Turbo message passing algorithm:}} First, the initialization of $\{\breve{s}_{n}\}$ and $\{\upsilon_{s_{n}}\}$ in Algorithm~\ref{alg.Turbo-Detection}
  is based on the PDF in \eqref{pro.s2} rather than in \eqref{pro.s}.
  Second, the variance of each $\tilde{w}_{\bsm{s},mt}$ in \eqref{var.s} is changed to
\begin{align}
\upsilon_{\tilde{w}_{\bsm{s},mt}}&=
\sum_{k=1}^K \upsilon_{x_{kt}} \left| \sum_{l=1}^{L}\left(\sum_{n=1}^{N_l} \rho_l h_{n,mk}\right) \right|^2 + \sigma_w^2 \notag\\
&+ \sum_{k'=1}^K  \sum_{l'=1}^{L}\left( \sum_{n'=1}^{N_l} \rho_l(1-\rho_l) \upsilon_{x_{k't}} \left|h_{n',mk} \right|^2 \right)  .
\end{align}
\end{itemize}

By the above modifications, we obtain the corresponding designs for the multi-RIS system.

\section{Numerical Results} \label{sec.Simul}

\subsection{Generation Model of the Channel} \label{sec.sub.channel}

We first describe the generation model of the channel used in simulations.
We model the channel by considering both small-scale and large-scale fading.
The direct link channel $\bsm{H}_0$ is modelled as follows. The line of sight (LoS) linkage of $\bsm{H}_0$ may be blocked while there usually exist a plenty of scatterers in the wireless channel. Thus, we follow \cite{wu2019intelligent,han2019large} to model the small-scale fading component of $\bsm{H}_0$ as Rayleigh fading with the elements independently taking over the CSCG distribution $\mathcal{CN}(0, 1)$.
The large-scale fading component of the direct link can be represented as $\diag\{\bsm{f}_0\}$, where $\bsm{f}_0=[f_{0,1},\ldots, f_{0,K}]^{\rm T}$ with $f_{0,k}$ being the larger-scale fading factor of the $k$th user.

We now describe the model of the reflect link channels $\bsm{G}_1$ and $\bsm{G}_2$.
The RIS is installed to ensure that the LoS linkages from the user to the RIS and from the RIS to the BS exist in the practical scenarios.
Thus, for the small-scale fading components of $\bsm{G}_1$ and $\bsm{G}_2$, we adopt a Rician fading model to account for both the LoS and non-LoS (NLoS) effects\cite{wu2019intelligent,han2019large}.
As explained in \cite{ntontin2019reconfigurable,basar2019wireless},  the RIS can be regarded as a mirror and produces only specular reflections. Thus, the large-scale fading channel of the reflect link can be represented as $\diag\{\bsm{f}_1\}$, where $\bsm{f}_1=[f_{1,1},\ldots, f_{1,K}]^{\rm T}$ with $f_{1,k}$ being the large-scale fading factor of user $k$.
The specific channel models of $\bsm{G}_1$ and $\bsm{G}_2$ are given by
\begin{align}
&\bsm{G}_1 = \left( \sqrt{\frac{ \kappa_{1} }{1 + \kappa_{1} } } \bar{\bsm{G}}_1 + \sqrt{\frac{ 1 }{1 + \kappa_{1} } } \tilde{\bsm{G}}_1 \right)\diag\{\bsm{f}_1\} \\
&\bsm{G}_2 = \left( \sqrt{\frac{ \kappa_{2} }{1 + \kappa_{2} } } \bar{\bsm{G}}_2 + \sqrt{\frac{ 1 }{1 + \kappa_{2} } } \tilde{\bsm{G}}_2 \right),
\end{align}
where  $\kappa_{1}$ and $\kappa_{2}$ are the corresponding Rician factors; $\bar{\bsm{G}}_1 = [\bar{\bsm{g}}_{1,1},\ldots,\bar{\bsm{g}}_{1,K}]$ and $\bar{\bsm{G}}_2$ denote the corresponding LoS components that remain fixed within a transmission block; $\tilde{\bsm{G}}_1$ and $\tilde{\bsm{G}}_2$  denote the corresponding NLoS components, where their elements are independently taken from the CSCG distribution $\mathcal{CN}(0, 1)$.

$\bar{\bsm{G}}_1$ and $\bar{\bsm{G}}_2$ can be further expressed as follows.
Assume that both the BS and the RIS are equipped with a two-dimensional ($2$D) uniform rectangular array.
The $2$D array steering vector $\mathbf{a}(\vartheta,\psi)$ is given by
\begin{align}
\mathbf{a}(\vartheta,\psi)&=\mathbf{a}_{\textsf{az}}(\vartheta,\psi)\otimes \mathbf{a}_{\textsf{el}}(\vartheta,\psi),
\end{align}
where $\vartheta$ and $\psi$ are the azimuth and elevation angles, respectively,
and $\mathbf{a}_{\textsf{az}}(\vartheta,\psi) \in \mathbb{C}^{M_1 \times 1}$ and $\mathbf{a}_{\textsf{el}}(\vartheta,\psi)\in \mathbb{C}^{M_2 \times 1}$ are the uniform linear array (ULA) steering vector given by
\begin{align}
\hspace{-0.2cm}\left[\mathbf{a}_{\textsf{az}}(\vartheta,\psi)\right]_n&= e^{-j \frac{2\pi d (n-1)}{\varrho}\sin(\vartheta) \cos(\psi)}, \forall n \in \mathcal{I}_{N_1} \\
\left[\mathbf{a}_{\textsf{el}}(\vartheta,\psi)\right]_n&= e^{j \frac{2\pi d (n-1)}{\varrho}\sin(\vartheta) \cos(\psi)}, \forall n \in \mathcal{I}_{N_2}
\end{align}
with $\varrho$ being the wavelength of propagation and $d$ being the distance of any two adjacent antennas.
Then, $\bar{\bsm{g}}_{1,k},\forall k \in \mathcal{I}_{K}$, and $\bar{\bsm{G}}_2$ can be expressed as
\begin{align}
\bar{\bsm{g}}_{1,k} &=\mathbf{a}(\vartheta_{1,k}^r,\psi_{1,k}^r), \forall k \in \mathcal{I}_{K} \\
 \bar{\bsm{G}}_{2}&=\mathbf{a}(\vartheta_{2}^r,\psi_{2}^r)\mathbf{a}^{\rm H}(\vartheta_{2}^t,\psi_{2}^t),
\end{align}
where $\mathbf{a}(\vartheta_{1,k}^r,\psi_{1,k}^r), \forall k \in \mathcal{I}_{K}$ is the arrival steering vector of the RIS for user $k$,
$\mathbf{a}(\vartheta_{2}^r,\psi_{2}^r)$ is the arrival steering vector of the BS, and
$\mathbf{a}(\vartheta_{2}^t,\psi_{2}^t)$ is the departure steering vector of the RIS.

In simulations,
$\kappa_1$ and $\kappa_2$ are set to $3$ dB and $10$ dB, respectively;  $\vartheta_{1,k}^r, \forall k$, $\vartheta_{2}^r$, and $\vartheta_{2}^t$ are randomly taken from a uniform distribution $[0, 2\pi)$;
$\psi_{1,k}^r, \forall k$,  $\psi_{2}^r$, and $\psi_{2}^t$ are randomly taken from a uniform distribution $[-\frac{\pi}{3}, \frac{\pi}{3}]$;
$\frac{d}{\varrho}=\frac{1}{2}$; the dimension of the $2$D array of the RIS in the single-RSI system is $16 \times \frac{N}{16}$; the dimension of the $2$D array of $\text{RIS}^{l}$ in the multi-RSI system is $16 \times \frac{N_l}{16}, \forall l \in \mathcal{I}_L$; the dimension of the $2$D array of the BS is $8 \times \frac{M}{8}$.
$\{f_{0,k}\}$ are independently taken from the uniform distribution $[-10, 0]$ dB.
$\{f_{1,k}\}$ are independently taken from the uniform distribution $[-13, 0]$ dB.


\subsection{Simulations for Passive Beamforming Design} \label{sec.sub.BF}
\begin{figure}[t]
  \centering
  \subfigure[$N = 32$]{\includegraphics[width=1.7 in]{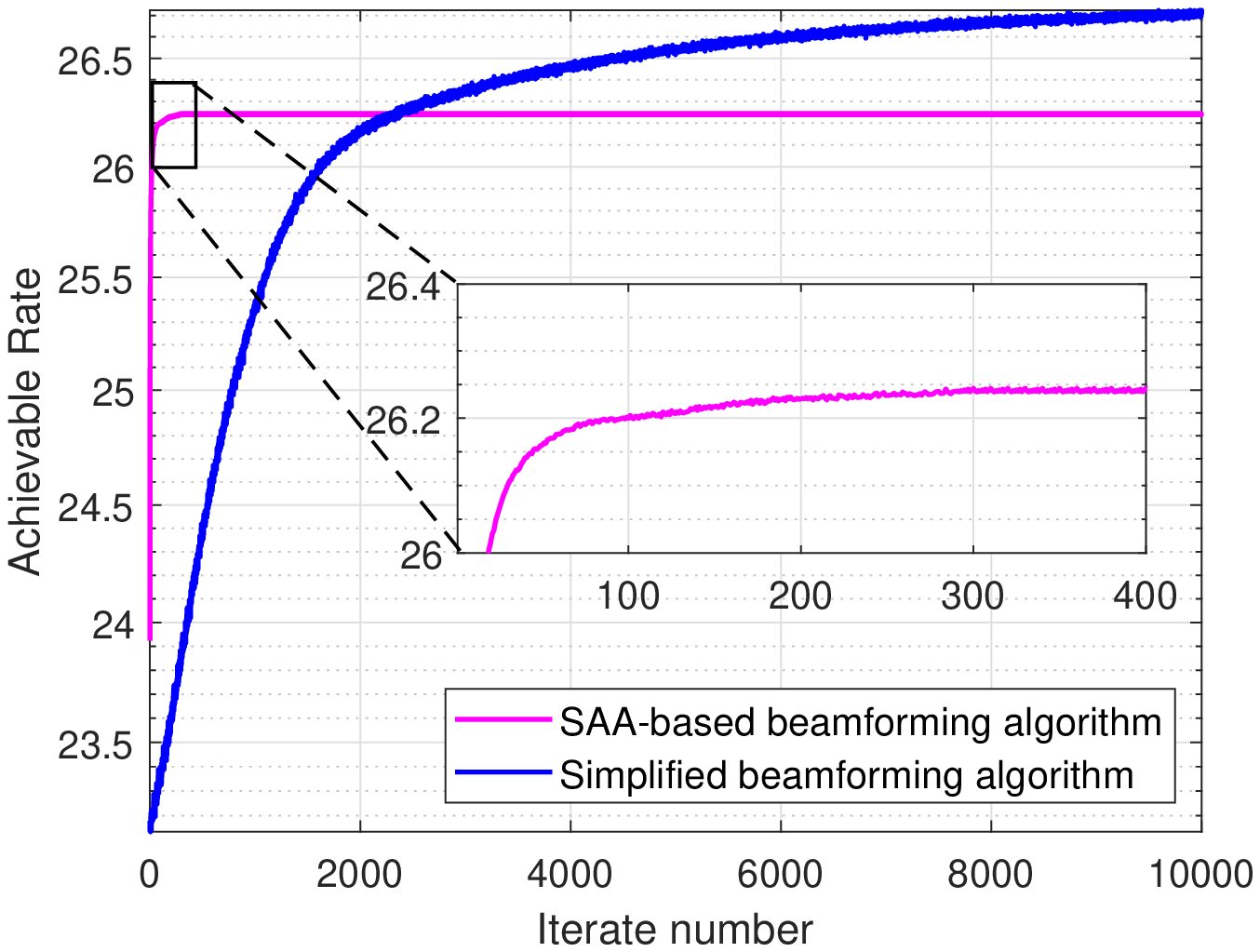}}
  \subfigure[$N = 64$]{\includegraphics[width=1.7 in]{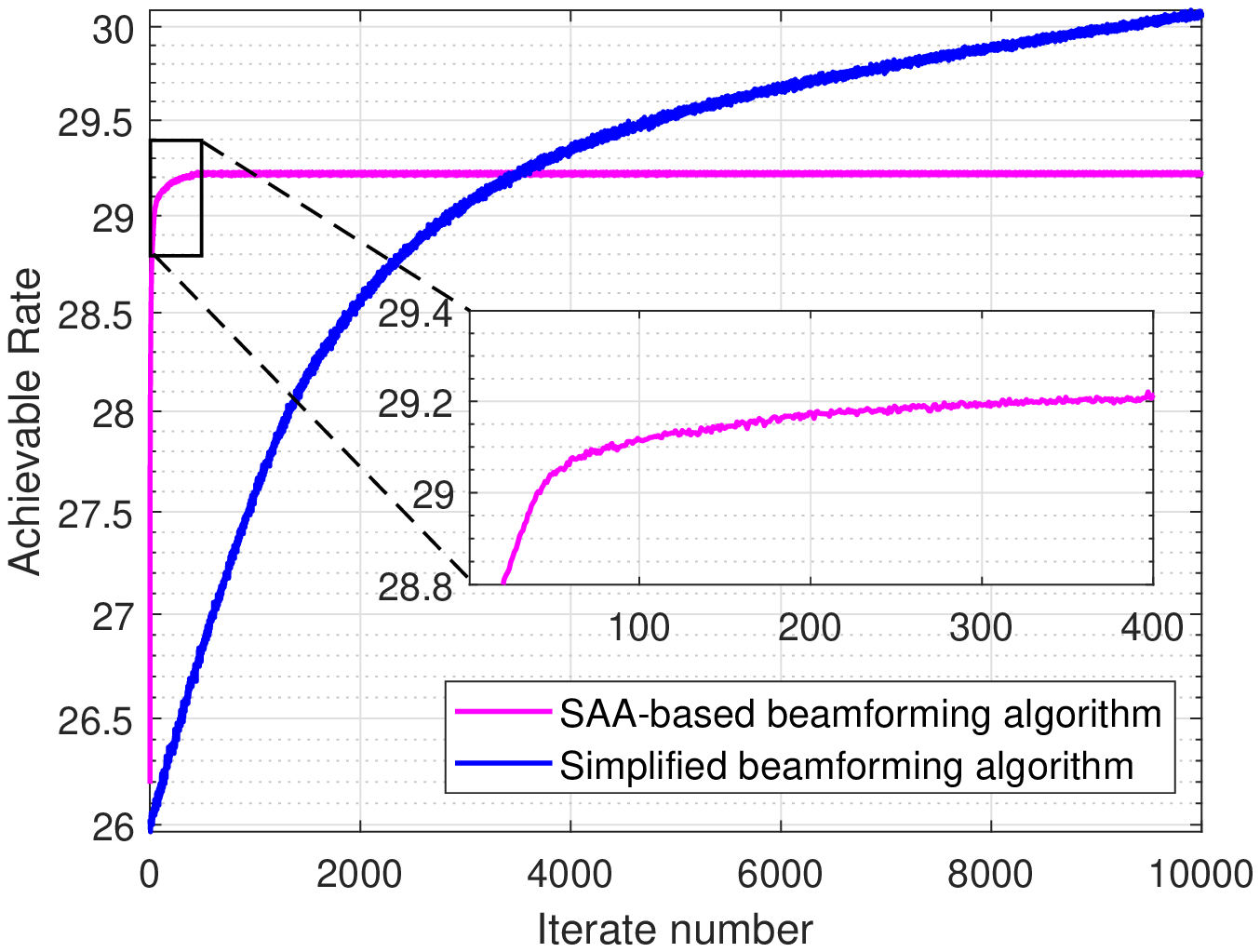}}
  \subfigure[$N = 96$]{\includegraphics[width=1.7 in]{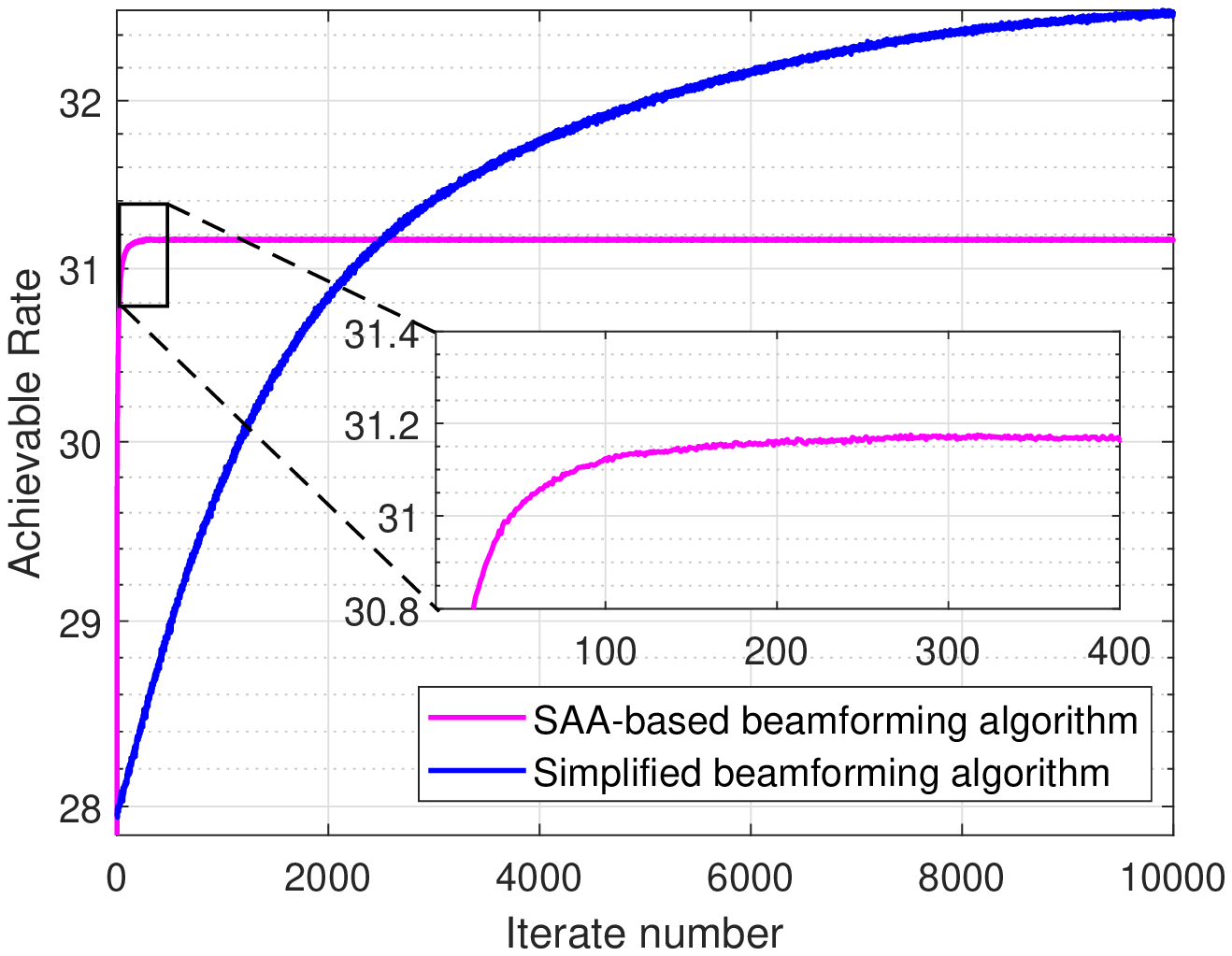}}
  \subfigure[$N = 128$]{\includegraphics[width=1.7 in]{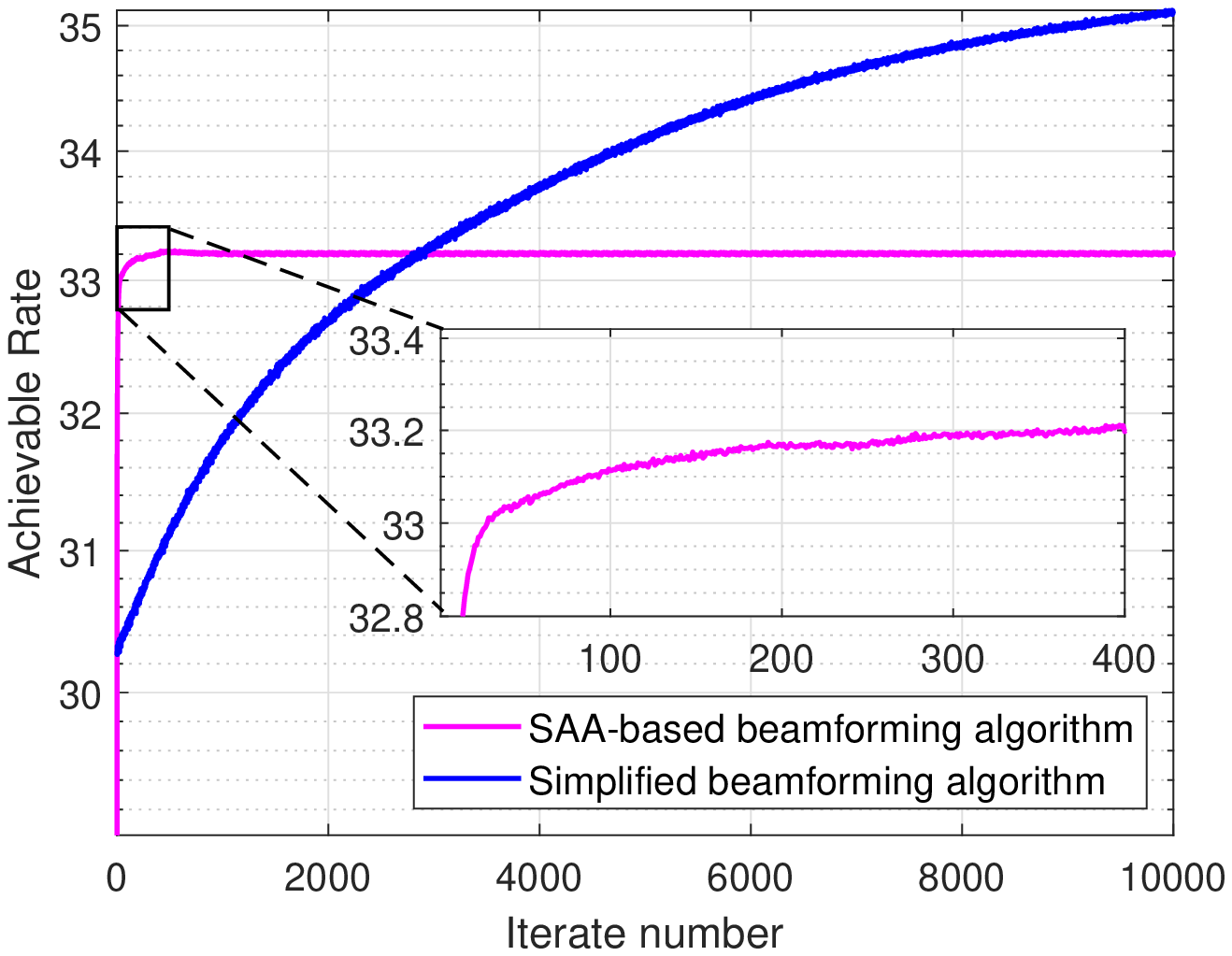}}
  \caption{Comparison of the achievable rates in the single-RIS system with the SAA beamforming algorithm and the simplified beamforming algorithm.}\label{IteraNum}
\end{figure}
\begin{figure}[t]
  \centering
  \includegraphics[width=3 in]{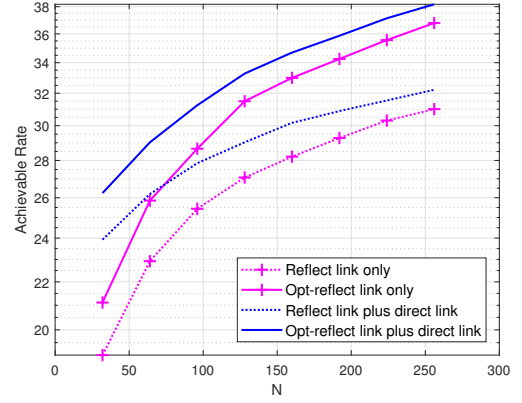}\\
  \caption{Achievable rate of the single-RIS system versus $N$.}\label{Rate_N}
\end{figure}
\begin{figure}[t]
  \centering
  \includegraphics[width=3 in]{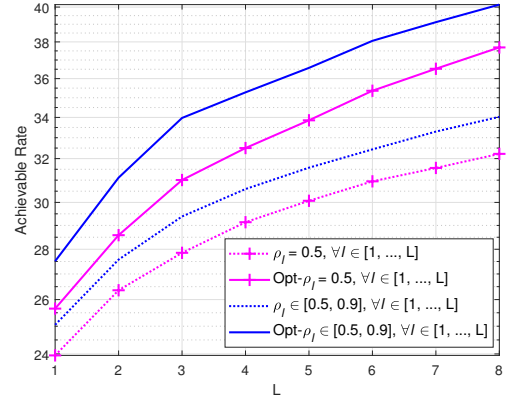}\\
  \caption{Achievable rate of the multi-RIS system versus $L$.}\label{Rate_L}
\end{figure}

In this subsection, we present numerical results to validate the efficiency of the proposed passive beamforming algorithms.
The parameter settings for the passive beamforming design are: $M=32$, $K=4$, $p_k=1,\forall k$, $\ell_{\bsm{s}}=100 $, $\ell_{\bsm{\theta}}=10 $, $\epsilon_{bf1}=10^{-4}$, $iter_{bf1}=10^4$, $\epsilon_{bf1}=10^{-2}$, and $iter_{bf2}=50$.
The presented simulation results are obtained by taking average over 100 random realizations.
The achievable rate of the system is calculated by using \eqref{optimal.Rate}.

Fig.~\ref{IteraNum} compares the achievable rates in the single-PBIT system versus the iteration number with the SAA-based beamforming algorithm and the simplified beamforming algorithm. We see that the SAA-based beamforming algorithm outperforms the simplified beamforming algorithm by about $0.5$ bit to $2$ bits with $N$ varying from $32$ to $128$ at the iteration number $=10^4$. We also see that the simplified beamforming algorithm achieves convergence at the iteration number $\approx200$ and is at least two orders of magnitude faster than the SAA-based beamforming algorithm. This demonstrates a good tradeoff between the performance and the computational cost.
 Since the computational cost of the SAA-based beamforming algorithm quickly becomes unaffordable as $N$ increases, in the remaining figures, we only present the simulation results of the simplified beamforming algorithm.

Fig.~\ref{Rate_N} studies the achievable rate of the single-PBIT system with $N$ varying from $32$ to $256$ with $\rho=0.5$.
We consider the following four scenarios:
$1)$ Reflect link only: the system only has the reflect link with random $\bsm{\theta}$ (i.e., $\bsm{\theta}$ is randomly taken from a uniform distribution $[0, 2\pi)$);
$2)$ Opt-reflect link only: the system only has the reflect link with optimized $\bsm{\theta}$ (i.e., $\bsm{\theta}$ is optimized by the simplified beamforming algorithm);
$3)$ Reflect link plus direct link: the system has both the reflect link and the direct link with random $\bsm{\theta}$;
$4)$ Opt-Reflect link plus direct link: the system has both the reflect link and the direct link with optimized $\bsm{\theta}$.
 We see that with $N$ increasing from $32$ to $256$, the gain obtained by the optimization of $\bsm{\theta}$ increases from $2$ bits to $6$ bits no matter the direct link exists or not. We also see that the rate gain contributed by the direct link decreases from $5$ bit to $1$ bit when $N$ is increased from $32$ to $256$.

Fig.~\ref{Rate_L} compares the achievable rate of the multi-RIS system with $L$ varying from $1$ to $8$.
We consider the following four scenarios:
$1)$ $\rho_l = 0.5, \forall l \in \mathcal{I}_L$ with random $\bsm{\theta}$ (uniform sparsity);
$2)$ $\rho_l = 0.5, \forall l \in \mathcal{I}_L$ with optimized $\bsm{\theta}$;
$3)$ $\rho_l, \forall l \in \mathcal{I}_L$ is randomly taken from a uniform distribution over $[0.5,0.9]$ with random $\bsm{\theta}$
(random sparsity);
$4)$ random sparsity $\rho_l, \forall l \in \mathcal{I}_L$ is randomly taken from a uniform distribution over $[0.5,0.9]$ with optimized $\bsm{\theta}$.
We see that with the increase of $L$, the gain obtained by the optimization of $\bsm{\theta}$ increase from $2$ bits to $6$ bits for both the uniform sparsity and random sparsity cases. We also see that the random sparsity cases outperform the uniform sparsity cases by $2$ bits throughout the considered range of $L$ no matter $\bsm{\theta}$ is random or maximized.

\subsection{Simulations for Detector Design} \label{sec.sub.De}

\begin{figure}[t]
  \centering
  \subfigure[Reflect link only]{\includegraphics[width=3.2 in]{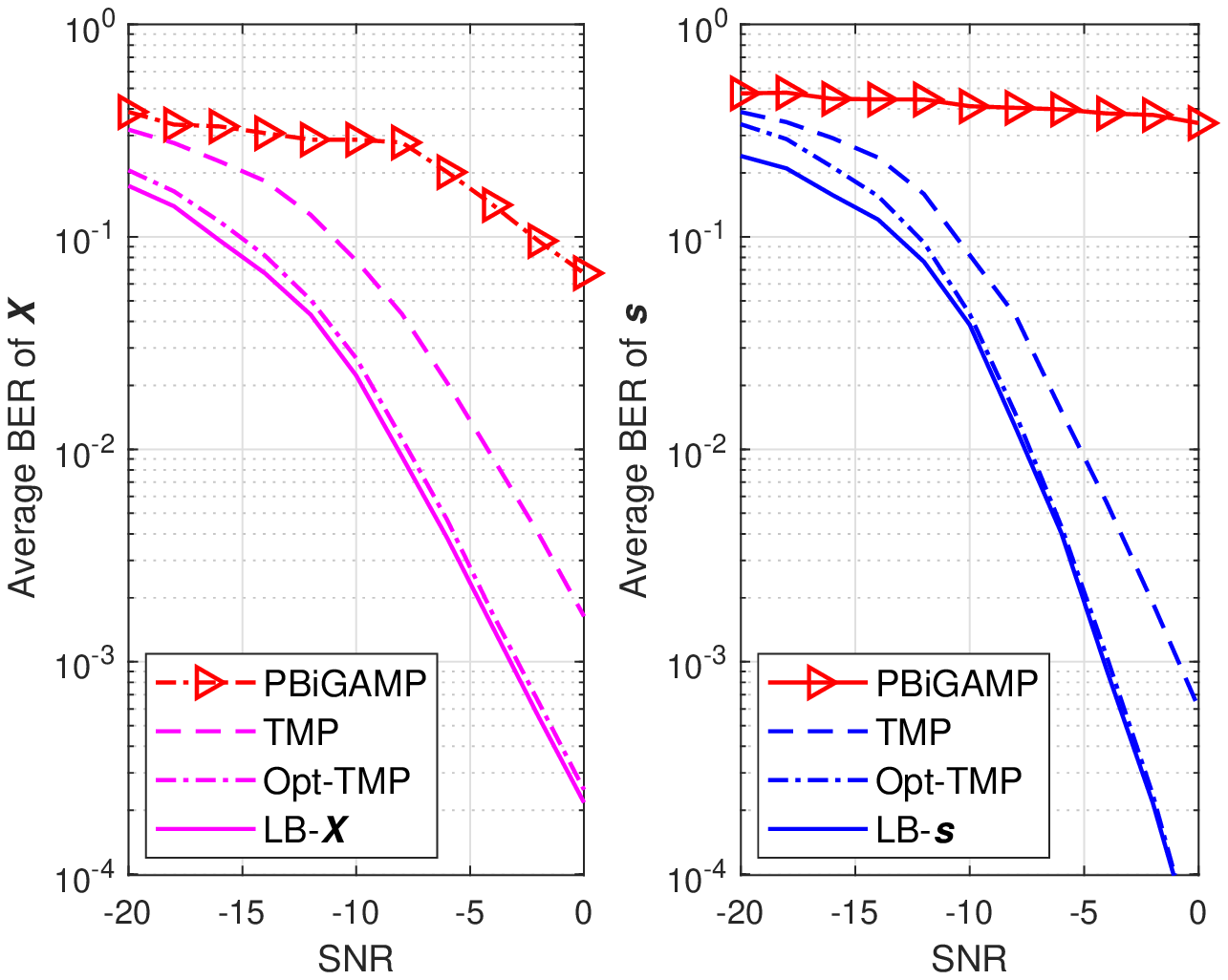}}
  \subfigure[Reflect link plus direct link]{\includegraphics[width=3.2 in]{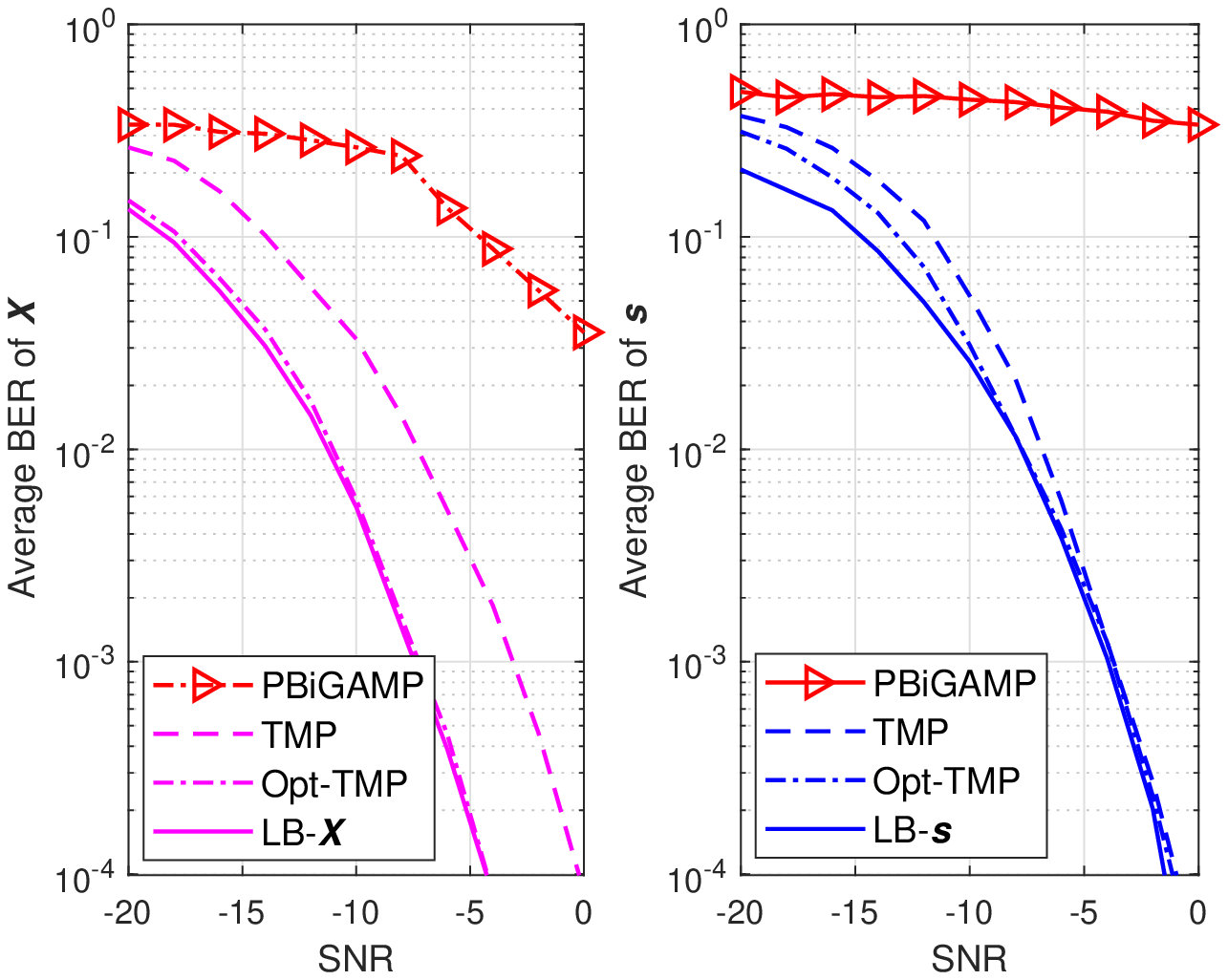}}
  \caption{The BERs of $\bsm{X}$ and $\bsm{s}$ of the single-RIS system versus the SNR with $N=128$, $M=32$, and $K=4$.} \label{Sig_XS}
\end{figure}

In this subsection, we present numerical results to validate the efficiency of the proposed the turbo message passing algorithm.
In simulations, the elements of $\bsm{X}$ are randomly taken from the quadrature phase shift keying modulation with Gray-mapping. We set $N_1=\ldots=N_L = 32$ in the multi-RIS cases.
The parameter settings for the $\bsm{X}$-detector are: $I_{max}^{x}=10$, $\ell_{max}^{x}=200$, $\xi_x^x =0.6$,
$\xi_u^x = \frac{2[(2-\xi_x^x)K + \xi_x^x M]}{1.1\xi_x^x MK \|\bsm{H}\|_2^2/\|\bsm{H}\|_{\rm F}^2} $,
 $ \epsilon_{\text{gamp}}^x = 10^{-10}$, and $\epsilon_{\text{em}}^x=10^{-10}$.
 The parameter settings for the $\bsm{s}$-detector are:
  $I_{max}^{s}=10$, $\ell_{max}^{s}=1000$, $\xi_x^s=0.2 $,
 $\xi_u^s  = \frac{2[(2-\xi_x^s)N + \xi_x^s MT]}{1.1\xi_x^s MTN \|\bsm{A}\|_2^2/\|\bsm{A}\|_{\rm F}^2} $,
 $ \epsilon_{\text{gamp}}^s= 10^{-8}$ and $\epsilon_{\text{em}}^s= 10^{-8}$.
The other settings are $\tau_{max}=20$, $\epsilon_{td}=10^{-3}$.
The presented simulation results are obtained by taking average over $500$ random realizations.

In simulations, we consider the following five scenarios:
$1)$ PBiGAMP: the receiver detects $\bsm{X}$ and $\bsm{s}$ by the PBiGAMP algorithm\cite{parker2016parametric} with optimized $\bsm{\theta}$;
$2)$ TMP: the receiver detects $\bsm{X}$ and $\bsm{s}$ by the turbo message passing algorithm proposed in this paper with random $\bsm{\theta}$;
$3)$ Opt-TMP: the receiver detects $\bsm{X}$ and $\bsm{s}$ by the turbo message passing algorithm proposed in this paper with optimized $\bsm{\theta}$;
$4)$ Lower bound for detecting $\bsm{X}$ (LB-$\bsm{X}$): the receiver detects $\bsm{X}$ by the GGAMP-SBL algorithm\cite{al2017gamp} with perfect knowledge of $\boldsymbol{s}$ and optimized $\bsm{\theta}$;
$5)$ Lower bound for detecting $\bsm{s}$ (LB-$\bsm{s}$) by the GGAMP-SBL algorithm\cite{al2017gamp} with perfect knowledge of $\boldsymbol{X}$ and optimized $\bsm{\theta}$.

Fig.~\ref{Sig_XS} compares the BERs of $\bsm{X}$ and $\bsm{s}$ of the single-RIS system versus the SNR with $N=128$, $M=32$, and $K=4$.
 We consider the PBiGAMP, TMP, Opt-TMP, LB-$\bsm{X}$, and LB-$\bsm{s}$ schemes for the cases of both with and without the direct link.
We see that the turbo message passing algorithm significantly outperforms the PBiGAMP algorithm and tightly approaches the lower bound no matter the direct link exists or not.
We also see that, by optimization, the BER of $\bsm{X}$ is improved by about $4$ dB throughout the considered SNR range, and the optimization of $\bsm{\theta}$ does not have much impact on the performance of the detecting $\bsm{s}$. The reason is that the target function of the optimization (i.e., $I(\bsm{x};\bsm{y}|\bsm{s})$) is actually the achievable rate of $\bsm{X}$.

\begin{figure}[t]
  \centering
  \subfigure[$\rho_l = 0.5, \forall l \in \mathcal{I}_L$]{\includegraphics[width=3.2 in]{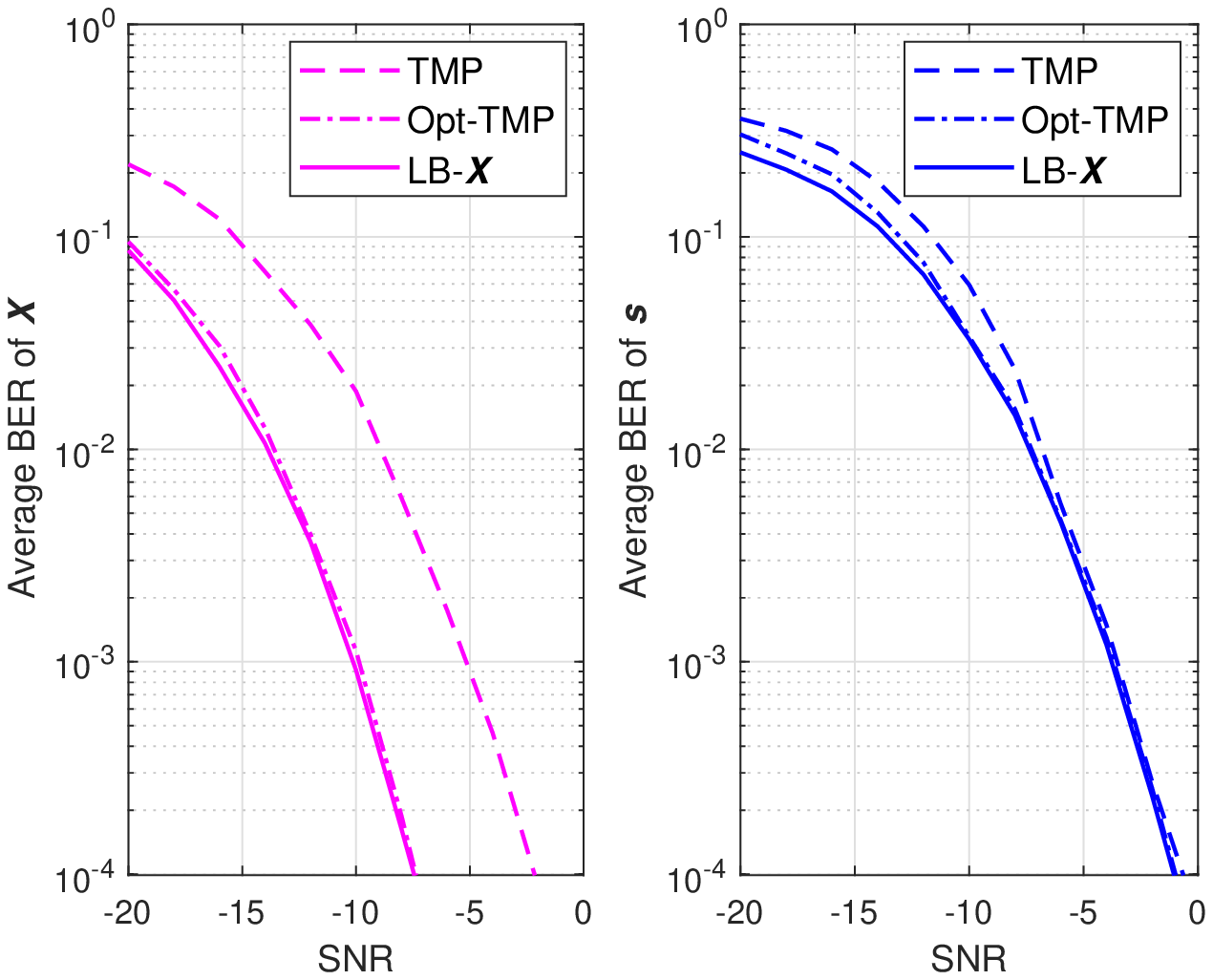}}
  \subfigure[$\rho_l$ randomly taken from {$[0.5,0.9]$},$\forall l \in \mathcal{I}_L$]{\includegraphics[width=3.2 in]{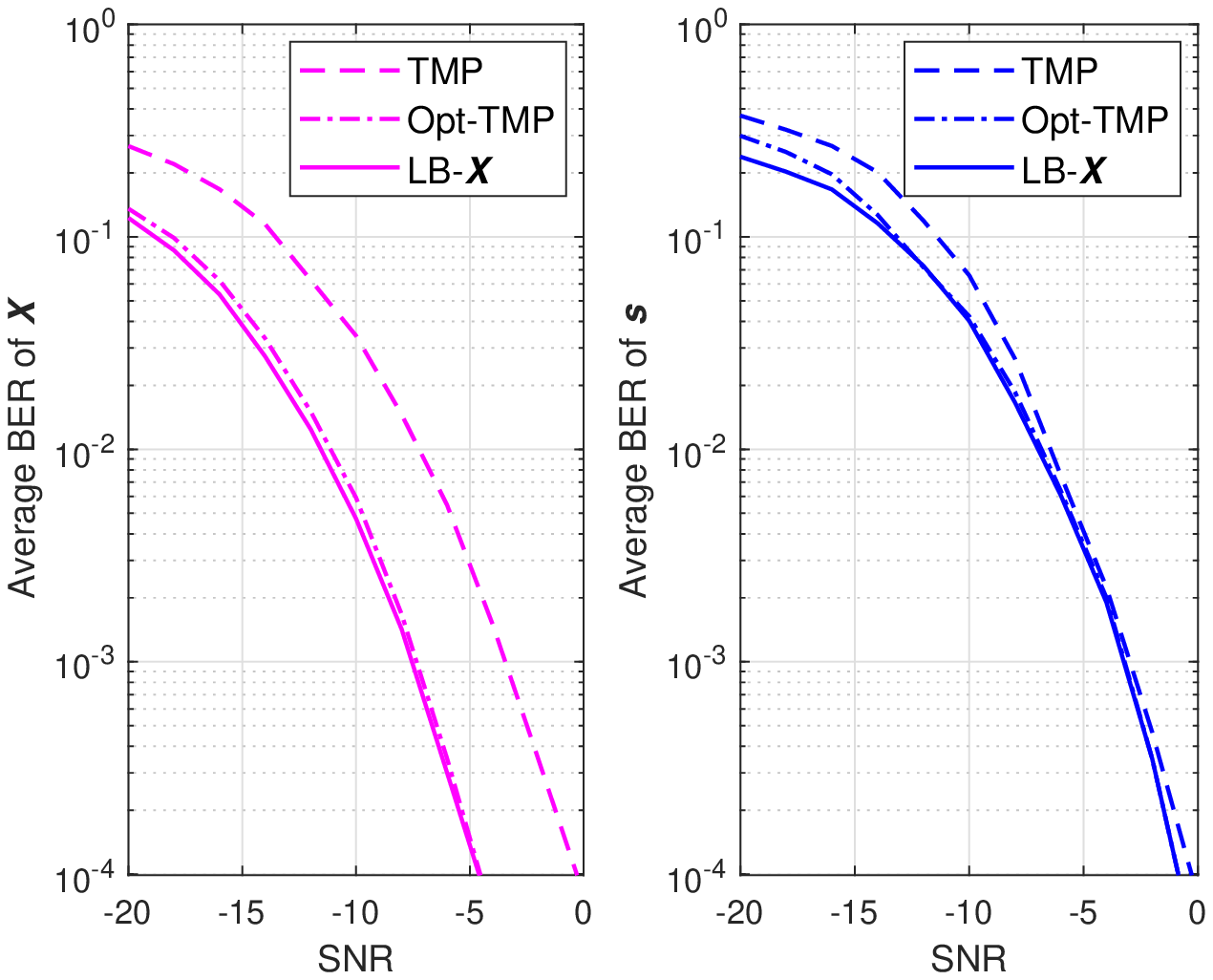}}
  \caption{The BERs of $\bsm{X}$ and $\bsm{s}$ of the multi-PBIT system versus SNR with $L=4$, $M=32$, $K=4$.} \label{Mul_XS}
\end{figure}
Fig.~\ref{Mul_XS} compares the BERs of $\bsm{X}$ and $\bsm{s}$ of the multi-PBIT system versus the SNR with $L=4$, $M=32$, and $K=4$.
 We consider the TMP, Opt-TMP, LB-$\bsm{X}$, and LB-$\bsm{s}$ schemes under uniform sparsity and random sparisty settings.
We see that, by optimization, the uniform sparsity scenario can achieve about $4$ dB performance improvement and the random sparsity scenario can achieve about $5$ dB performance improvement at the average BER $=10^{-4}$.
We also see that the turbo message passing algorithm tightly approaches the lower bounds for both the uniform and random sparsity cases.

\section{Conclusions}
\label{sec.Con}

In this paper, we studied the design of passive beamforming and information transfer for the RIS-aided multiuser MIMO system. For the passive beamforming design, we formulate the problem as a stochastic program in which the conditional mutual information of the RIS-aided multiuser MIMO channel is used as the metric for optimization, and developed the SAA-based beamforming algorithm to solve the problem. To reduce the high computational complexity of the SAA-based beamforming algorithm, we further developed a simplified beamforming algorithm by approximating the stochastic program as a deterministic alternating optimization problem. For the receiver design, we proposed a turbo message passing algorithm to iteratively estimate the user signals and the RIS states. Furthermore, we extended all the designs from the single-RIS case to the multi-RIS case. Numerical results were provided to demonstrate the superior performance of our proposed designs.

\section*{Appendix A}  \label{Appendix.A}

From the Blahut-Arimoto algorithm in \cite{cover2012elements},
$I(\bsm{x};\bsm{y}|\bsm{s})$ can be written as
\begin{subequations} \label{Inf.1}
\begin{align}
I(\bsm{x};\bsm{y}|\bsm{s}) &=  \mbs{E}_{\bsm{y},\bsm{x},\bsm{s}}
\log \frac{p_{\bsm{x},\bsm{y}|\bsm{s}}(\bsm{x},\bsm{y}|\bsm{s})}
{p_{\bsm{x}|\bsm{s}}(\bsm{x}|\bsm{s})p_{\bsm{y}|\bsm{s}}(\bsm{y}|\bsm{s})}   \label{Inf.1.a}\\
&=  \mbs{E}_{\bsm{y},\bsm{x},\bsm{s}}
\log \frac{p_{\bsm{x}|\bsm{y},\bsm{s}}(\bsm{x}|\bsm{y},\bsm{s})}{p_{\bsm{x}}(\bsm{x})}  \label{Inf.1.b}\\
&=  \max_{\tilde{p}(\cdot | \cdot)} \mbs{E}_{\bsm{y},\bsm{x},\bsm{s}} \log \frac{\tilde{p}(\bsm{x}|\bsm{y},\bsm{s})}{p_{\bsm{x}}(\bsm{x})} ,\label{Inf.1.c}
\end{align}
\end{subequations}
where the expectation $\mbs{E}$ is taken over the joint PDF of $\bsm{y}$, $\bsm{x}$ and $\bsm{s}$,  \eqref{Inf.1.b} is based on the fact that $\bsm{x}$ and $\bsm{s}$ are independent of each other, and $\tilde{p}(\cdot | \cdot)$ in \eqref{Inf.1.c} is an arbitrary distribution of $\bsm{x}$ conditioned on $\bsm{y}$.

Define an auxiliary variable $\bsm{z} = ( \bsm{G}_2 \bsm{\Theta}\bsm{S}\bsm{G}_1  + \bsm{H}_0 )\bsm{x}$ and recall that $\bsm{w}$ is an AWGN with the elements independently drawn from $\mathcal{CN}(0,\sigma_w^2)$. Then, we have
\begin{align}
p_{\bsm{y}|\bsm{x},\bsm{s}}(\bsm{y}|\bsm{x},\bsm{s}) = \frac{1}{(\pi \sigma_w^2 )^M } \exp(- \frac{1}{\sigma_w^2} \left(\bsm{y}- \bsm{z}\right)^{\rm H} \left(\bsm{y}- \bsm{z}\right) ).
\end{align}
Since $ I(\bsm{x};\bsm{y}|\bsm{s})$ is maximized when $\bsm{x}$ is Gaussian,
 we consider the Gaussian input:
\begin{align} \label{pro.x2}
p_{\bsm{x}}(\bsm{x}) = \frac{1}{\pi^K \det(\bsm{Q})} \exp(- \bsm{x}^{\rm H}\bsm{Q}^{-1}\bsm{x} ),
\end{align}
where $\bsm{Q} = \mbs{E}\left(\bsm{x}\bsm{x}^{\rm H} \right)$ is the covariance matrix of $\bsm{x}$.
Note that $\bsm{Q}$ is a diagonal matrix and the $k$th diagonal element is given by $p_k, \forall k \in \mathcal{I}_{K} $. The probability distribution of $\bsm{s}$ is given in \eqref{pro.s}.
Then, the joint PDF $p_{\bsm{y},\bsm{x},\bsm{s}}(\bsm{y},\bsm{x},\bsm{s})$ is given by
\begin{align}
& p_{\bsm{y},\bsm{x},\bsm{s}}(\bsm{y},\bsm{x},\bsm{s}) \notag\\
&= p_{\bsm{y}|\bsm{x},\bsm{s}}(\bsm{y}|\bsm{x},\bsm{s}) p_{\bsm{x}}(\bsm{x}) p_{\bsm{s}}(\bsm{s}) \notag\\
&= \frac{1}{(\pi \sigma_w^2 )^M } \exp(- \frac{1}{\sigma_w^2} \left(\bsm{y}- \bsm{z}\right)^{\rm H} \left(\bsm{y}- \bsm{z}\right) ) \notag\\
&~~~~ \times \frac{1}{\pi^K \det(\bsm{Q})} \exp(- \bsm{x}^{\rm H}\bsm{Q}^{-1}\bsm{x} ) \notag\\
&~~~~\times  \prod_{n=1}^N \left[ (1-\rho)\delta(s_n) + \rho \delta(s_n-1) \right].
\end{align}

From \cite{kay1993fundamentals},
the optimal choice of $\tilde{p}(\cdot | \cdot)$ follows the Gaussian distribution and can be written as
\begin{align} \label{pro.x3}
&p_{\bsm{x}|\bsm{y},\bsm{s}}(\bsm{x}|\bsm{y},\bsm{s}) \notag\\
&= \frac{1}{\pi^K \det(\bsm{\Sigma})} \exp(- \left(\bsm{x}- \bsm{\Phi}\bsm{y}\right)^{\rm H}\bsm{\Sigma}^{-1}\left(\bsm{x}- \bsm{\Phi}\bsm{y}\right) ),
\end{align}
where $\bsm{\Phi}$ is a coefficient matrix to be determined, $\bsm{\Phi}\bsm{y}$ represents
the conditional mean, and $\bsm{\Sigma}$ represents the conditional
variance.
Plugging \eqref{pro.x2} and \eqref{pro.x3} into \eqref{Inf.1.c}, we obtain
\begin{align}  \label{optimal.Rate}
&I(\bsm{x};\bsm{y}|\bsm{s}) = \mbs{E}_{\bsm{s}}\left[ \max_{\bsm{\Phi},\bsm{\Sigma}} \mbs{E}_{\bsm{y},\bsm{x}|\bsm{s}}\left( \log\det(\bsm{Q})-\log\det(\bsm{\Sigma}) \right.\right. \notag\\
&~~~~~~~~~~~~- \left.\left.\left(\bsm{x}- \bsm{\Phi}\bsm{y}\right)^{\rm H}\bsm{\Sigma}^{-1}\left(\bsm{x}- \bsm{\Phi}\bsm{y}\right) + \bsm{x}^{\rm H}\bsm{Q}^{-1}\bsm{x} \right)\right].
\end{align}
Then, we can obtain \eqref{optimal.e} by
omiting the terms $\mbs{E}_{\bsm{s}}\left[ \mbs{E}_{\bsm{y},\bsm{x}|\bsm{s}} \log\det(\bsm{Q})\right]$ and
$ \mbs{E}_{\bsm{s}}\left[\mbs{E}_{\bsm{y},\bsm{x}|\bsm{s}}\bsm{x}^{\rm H}\bsm{Q}^{-1}\bsm{x} \right]$ since they two are irrelevant to $\bsm{\Phi}$, $\bsm{\Sigma}$, and $\bsm{\Theta}$.

\section*{Appendix B} \label{Appendix.B}

From the model in \eqref{channel.H2}, we obtain
\begin{align} \label{optimal.b3}
&\mbs{E}_{\bsm{y},\bsm{x},\bsm{s}} \left\|\bsm{\Sigma}^{-\frac{1}{2}}\left(\bsm{x}- \bsm{\Phi}\bsm{y}\right)\right\|_2^2  \notag\\
&= \tr \left\{ \bsm{\Sigma}^{-1} \left[ \bsm{Q} - \bsm{\Phi} \bsm{H}_0 \bsm{Q}  - (\bsm{\Phi} \bsm{H}_0 \bsm{Q})^{\rm H} + \bsm{\Phi}\bsm{H}_0 \bsm{Q} \bsm{H}_0^{\rm H}\bsm{\Phi}^{\rm H}  \right.\right. \notag\\
&~~~~ + \sigma_{w}^2 \bsm{\Phi}\bsm{\Phi}^{\rm H} + \mbs{E}_{\bsm{s}} \left(\bsm{\Phi} \bsm{G}_2 \bsm{\Theta}\bsm{S}\bsm{G}_1 \bsm{Q} \bsm{G}_1^{\rm H} \bsm{S}^{\rm H} \bsm{\Theta}^{\rm H} \bsm{G}_2^{\rm H} \bsm{\Phi}^{\rm H} \right.  \notag\\
&~~~~ + \bsm{\Phi}\bsm{G}_2 \bsm{\Theta}\bsm{S}\bsm{G}_1 \bsm{Q} \bsm{H}_0^{\rm H}\bsm{\Phi}^{\rm H} + (\bsm{\Phi}\bsm{G}_2 \bsm{\Theta}\bsm{S}\bsm{G}_1 \bsm{Q} \bsm{H}_0^{\rm H}\bsm{\Phi}^{\rm H})^{\rm H}  \notag\\
&~~~~ \left.\left.\left.   -\bsm{\Phi} \bsm{G}_2 \bsm{\Theta}\bsm{S}\bsm{G}_1 \bsm{Q} - (\bsm{\Phi} \bsm{G}_2 \bsm{\Theta}\bsm{S}\bsm{G}_1 \bsm{Q})^{\rm H}
\right)\right]\right\}.
\end{align}
Based on the probability distribution of $\bsm{s}$ in \eqref{pro.s}, we obtain
\begin{align}
& \mbs{E}_{\bsm{s}}\left(\bsm{S}\right) = \rho \bsm{I}  \label{optimal.b4} \\
& \mbs{E}_{\bsm{s}}\left( \bsm{S}\bsm{X}\bsm{S}^{\rm H} \right) =\rho^2 \bsm{X} + \rho(1-\rho)\left[\bsm{X} \right]_{\diag}.
\label{optimal.b5}
\end{align}
Plugging \eqref{optimal.a10}, \eqref{optimal.b4}, and \eqref{optimal.b5} into \eqref{optimal.b3}, we obtain \eqref{optimal.b7}.

\section*{Appendix C} \label{Appendix.B}
\subsection{Derivation of \eqref{variance.x}}
Denote by $\bsm{h}_{n,m}^{\rm T}$ the $m$th row of $\bsm{H}_n$ and let
 $ \bsm{B}_m \triangleq [\bsm{h}_{0,m}, \ldots, \bsm{h}_{N,m}]^{\rm T} \in \mathbb{C}^{(N+1)\times K},\forall m\in \mathcal{I}_M $.
Then,  each element of $\tilde{\bsm{W}}_{\bsm{X}}$ can be written as $\tilde{w}_{\bsm{X},mt} = \left(\tilde{\bsm{s}}\right)^{\rm T}\bsm{B}_m\bsm{x}_t + w_{mt}, \forall m,t$. The variance of
$\tilde{w}_{\bsm{X},mt}$ is
\begin{align}
\upsilon_{\tilde{w}_{\bsm{X},mt}} &=  \mbs{E}\left[\left(\tilde{\bsm{s}}^{\rm T}\bsm{B}_{m}\bsm{x}_t + w_{mt}\right)\left(\tilde{\bsm{s}}^{\rm T}\bsm{B}_{m}\bsm{x}_t + w_{mt}\right)^{\rm H} \right] \notag\\
&= \mbs{E} \left( \tilde{\bsm{s}}^{\rm T} \bsm{B}_{m}\bsm{Q} \bsm{B}_{m}^{\rm H} \tilde{\bsm{s}}\right)  + \sigma_w^2 \notag\\
&= \sum_{i=0}^N\sum_{j=0}^N \mbs{E}\left(  \tilde{s}_i  \tilde{s}_j  \right) \left(\bsm{h}_{i,m}\bsm{Q}\bsm{h}_{j,m}^{\rm H}\right) + \sigma_w^2 \notag\\
&= \sum_{i=0}^N \sum_{k=1}^K \upsilon_{s_i}  |h_{i,mk}|^2 p_k + \sigma_w^2 ,\forall m,t,
\end{align}
where $ \mbs{E}\left(  \tilde{s}_i  \tilde{s}_j  \right) = \upsilon_{s_i}$ for $i=j \in \mathcal{I}_{N}$ and $ \mbs{E}\left(  \tilde{s}_i  \tilde{s}_j  \right) = 0$ otherwise.

\subsection{Derivation of \eqref{variance.s}}
\vspace{-0.5cm}

\begin{align} \label{var.s}
\upsilon_{\tilde{w}_{\bsm{s},mt}}&= \mbs{E}\left[\left(\bsm{s}^{\rm T}\bsm{B}_{m}\tilde{\bsm{x}}_t + w_{mt}\right)^{\rm H} \left(\bsm{s}^{\rm T}
\bsm{B}_{m}\tilde{\bsm{x}}_t + w_{mt}\right) \right] \notag\\
&= \mbs{E}\left[\tilde{\bsm{x}}_t^{\rm H} \bsm{B}_{m}^{\rm H} \left(\rho^2 \textbf{1}_N\cdot \textbf{1}_N^{\rm T} + \rho(1-\rho)\mathbf{I}\right) \bsm{B}_{m} \tilde{\bsm{x}}_t \right]  + \sigma_w^2 \notag\\
&= \rho^2 \mbs{E}\left[\tilde{\bsm{x}}_t^{\rm H} \left(\textbf{1}_N^{\rm T}\bsm{B}_{m}\right)^{\rm H} \left(\textbf{1}_N^{\rm T}\bsm{B}_{m} \right) \tilde{\bsm{x}}_t \right] \notag\\
&~~~~+ \rho(1-\rho)\mbs{E}\left[\tilde{\bsm{x}}_t^{\rm H} \bsm{B}_{m}^{\rm H} \bsm{B}_{m} \tilde{\bsm{x}}_t \right]+ \sigma_w^2   \notag\\
&= \rho^2\sum_{i=1}^K\sum_{j=1}^K \mbs{E}\left(\tilde{x}_{it}^{\ast} \tilde{x}_{jt}\right) \left(\textbf{1}_N^{\rm T}\bsm{B}_{m}\right)_i^{\ast} \left(\textbf{1}_N^{\rm T}\bsm{B}_{m}\right)_j \notag\\
&~~~~+ \rho(1-\rho) \sum_{i'=1}^K\sum_{j'=1}^K \mbs{E}\left(\tilde{x}_{i't}^{\ast} \tilde{x}_{j't}\right) \bsm{b}_{m,i'}^{\rm H}\bsm{b}_{m,j'} + \sigma_w^2  \notag\\
&=  \rho(1-\rho)\sum_{k'=1}^K \sum_{n'=1}^N \upsilon_{x_{k't}} \left|h_{n',mk} \right|^2 \notag\\
&~~~~ + \rho^2 \sum_{k=1}^K \upsilon_{x_{kt}} \left| \sum_{n=1}^N h_{n,mk} \right|^2  + \sigma_w^2 ,
\end{align}
where $\mbs{E}\left[\boldsymbol{s} \boldsymbol{s}^{\rm T}\right] = \rho^2 \textbf{1}_N\cdot \textbf{1}_N^{\rm T} + \rho(1-\rho)\mathbf{I} $,
  $\mbs{E}\left( \tilde{x}_{it}^{\ast} \tilde{x}_{jt} \right)=\upsilon_{x_{it}}$ for $i=j \in \mathcal{I}_{K}$ and $\mbs{E}\left( \tilde{x}_{it}^{\ast} \tilde{x}_{jt} \right)=0$ otherwise,
and $\bsm{b}_{m,k}$ is the $k$th column of $\bsm{B}_{m}$.

\bibliographystyle{IEEEtran}
\bibliography{MIMO}

\end{document}